\documentclass[reprint,aps,twocolumn,showpacs,amsmath,amssymb,pre]{revtex4-1}

\usepackage{graphicx}
\usepackage{textcomp}
\usepackage{psfrag}
\usepackage[colorlinks,linkcolor={blue},citecolor={blue},urlcolor={blue}]{hyperref}
\usepackage{dcolumn}
\usepackage{bm}
\usepackage{multirow}
\usepackage{array}
\usepackage{booktabs}
\usepackage{ctable}
\usepackage{upgreek}
\usepackage{epsfig}
\usepackage{mathrsfs}
\usepackage{amssymb}
\usepackage{amsbsy}
\usepackage{color}
\usepackage{cancel}
\usepackage{pifont}
\usepackage{marginnote}

\begin{document}
\title{Projecting Low Dimensional Chaos from  Spatio-temporal Dynamics in a Model for Plastic Instability}
\author{Ritupan Sarmah}\email{ritupan@mrc.iisc.ernet.in}
\author{G. Ananthakrishna}\email {garani@mrc.iisc.ernet.in}
\affiliation{Materials Research Centre, Indian Institute of Science, Bangalore 560012, India}

\begin{abstract}
We  investigate the possibility  of projecting low dimensional chaos from  spatiotemporal dynamics  of a model  for  a kind of plastic instability observed under constant strain rate deformation conditions. We first discuss the relationship between the spatiotemporal patterns of the model reflected in the nature of dislocation bands and the nature of stress serrations. We show that at low applied strain rates, there is a one-to-one correspondence with the randomly nucleated isolated bursts of mobile dislocation density and the stress drops.  We then show that the model equations  are spatiotemporally chaotic by demonstrating  the number of  positive Lyapunov exponents  and Lyapunov dimension scale with the system size at low and high strain rates.  Using a modified algorithm for calculating correlation dimension density, we show that the stress-strain signals at  low applied strain rates corresponding to spatially uncorrelated dislocation bands exhibit features of low dimensional chaos. This is made quantitative by demonstrating that the model equations can be approximately reduced to space independent model equations for the average dislocation densities, which is known to be low-dimensionally chaotic. However, the scaling regime for the correlation dimension shrinks with increasing applied strain rate due to increasing propensity for  propagation of the dislocation bands. 
\end{abstract}
 

\maketitle

\section{Introduction}

It is well known that a number of irregular scalar experimental signals have features of  low dimensional (low-d) chaos. These signals often correspond to a spatial average over internal degrees of freedom of a spatially extended system.  In such cases,  while the  internal degrees of freedom are not accessible to experiments, they are suspected  to be spatio-temporally chaotic, a view supported by  models that capture the basic features of the phenomenon.  A good example is the plastic deformation of metallic alloys subjected to a constant strain rate test. Here,  only stress can be measured by  the load cell placed at one end of the sample and is the spatial average of dislocation activity in the sample. Under specific conditions, the measured stress-time series, $\sigma(t)$, exhibits irregular serrations \cite{Cottrell53,GA07}. Such stress signals  have been shown to be low-d chaotic \cite{Noro97,GA99,Bhar01}. A second example is  the crackling audible noise, called acoustic emission (AE) commonly experienced during peeling of an adhesive tape. The AE signals have also been shown to be low-d chaotic \cite{Rumiprl}, which is controlled by the rugged nature of the peel front. These two examples suggest that spatial averaging  some how projects low-dimensional chaos out of spatio-temporal dynamics.  
Several candidates for low-d chaos could be cited, including  the  irregular voltage fluctuations measured across a sample of a charge density wave compound (CDW), attributed to pinning and unpinning of CDW \cite{Dumas}, and the fluctuations in the position and intensity of a light beam passing through a turbulent medium \cite{Dimo01}. 

The purpose of this paper is to examine the possibility of projecting low-d chaos from a spatiotemporal chaotic system and if so, under what conditions? This question will be addressed within the context of a model for a plastic instability, called  the Portevin-Le Chatelier (PLC) effect \cite{Anan82,Bhar02,Bhar03a,Bhar03b}. The PLC effect or jerky flow refers to the irregular  stress-strain curves associated with the observed heterogeneous deformation when specimens of dilute metallic alloys are deformed in a  window of strain rates and temperatures.  While  dislocation activity in the sample is not accessible to experiments, dislocation bands seen on  the surface of the sample \cite{GA07,HN83}, measurement of acoustic emission studies \cite{Weiss03}, and studies on micro-crystals \cite{Dimiduk06}, strongly suggest that dislocation dynamics is intermittent. However, relating the irregular nature of the scalar signal to the dislocation dynamics has remained a difficult task due to lack of dislocation based models even though there is a general consensus that dislocation dynamics is intermittent, a view supported by dislocation dynamics simulations as well \cite{Kubin08,Csikor07}.

The basic mechanism attributed to the stress serrations in the PLC effect is the collective pinning and unpinning of dislocations \cite{Cottrell53,GA07}.  Each stress drop is associated with the formation and often the propagation of dislocation bands. Three distinct types of dislocation bands, namely, the static uncorrelated type C, hopping type B and continuously propagating type A are seen with increasing strain rate. The average stress drop size is large for the type C band, decreasing with increasing strain rate as we encounter the type B and A bands \cite{GA07}.

Several of these generic features of the PLC effect are captured by the  Ananthakrishna (AK) model \cite{GA07,Anan82,Bhar02,Bhar03a,Bhar03b}. One prediction specific to the model relevant for the current study is the (low-d) chaotic nature of the stress drops for low strain rates \cite{Anan83}. The prediction has been subsequently verified using experimental signals from single and polycrystals \cite{Noro97,GA99,Bhar01}. Thus, the model provides a platform to  verify the conjuncture that the stress signals are low-d chaotic while the model equations are spatiotemporally chaotic. Using the AK model, we show that at low strain rates, the irregular stress-time series can be unambiguously identified as low-d chaotic while the model equations are spatiotemporally chaotic.  We show that the low-d chaotic nature is entirely due to the one-to-one correspondence between the isolated bursts of mobile density and the stress drops in this domain of applied strain rates. The low-d chaotic feature  breaks down for higher applied strain  rates  which can be directly correlated with the propensity for the dislocation bands to propagate.

\section{The Ananthakrishna Model for Serrated Flow}

The basic idea of the AK model is that most generic features of the PLC effect such as the existence of the instability in a window of strain rates, the negative strain rate sensitivity of the flow stress, and the three types of bands (C, B, and A) emerge from the nonlinear interaction of a few dislocation populations assumed to represent the collective degrees of freedom of the system \cite{Anan82,Bhar02,Bhar03a,Bhar03b}. The model uses three types of dislocation densities: the fast mobile $\rho_m (x,t)$, the immobile $\rho_{im} (x,t) $, and the Cottrell type $\rho_c (x,t) $ corresponding to dislocations decorated by solute atoms \cite{Anan82}. (The standard AK model equations take into account the hardening process that is relevant for describing  the physical phenomenon \cite{Anan82,Bhar02,Bhar03a,Bhar03b}. However,  since we are concerned here with the spatiotemporal dynamics in a stationary situation, we ignore  the hardening term that also helps to attain the stationary state faster.)  The scaled equations for the three densities and the scaled stress $\phi$  are given by 
\begin{eqnarray}
\nonumber
\frac{\partial \rho_m}{\partial t} &=& -b_0 \rho_m^2 -\rho_m\rho_{im} + \rho_{im} -a\rho_m + \phi^m \rho_m \\
\label{X-eqn}
&+& \frac{D\phi^m(t)}{\rho_{im}} \frac{\partial^2  \rho_m}{\partial x^2}, \\
\label{Y-eqn}
\frac{\partial \rho_{im}}{\partial t} &=& b_0(b_0\rho_m^2 -\rho_m\rho_{im} -\rho_{im} + a\rho_c),\\
\label{Z-eqn}
\frac{\partial \rho_c}{\partial t} &= &c(\rho_m - \rho_c ),\\
\label{phi-eqn}
\frac{d\phi(t)}{dt}& = &d[\dot{\epsilon}_a -\frac{\phi^m(t)}{l}\int_0^l \rho_m(x,t) dx] = d[\dot{\epsilon}_a -\dot{\epsilon}_p].
\end{eqnarray}
The term $b_0\rho_m^2$ in Eq. (\ref{X-eqn}),  refers to the formation of dipoles and other  dislocations locks,  $\rho_m\rho_{im}$  refers to the annihilation of a mobile dislocation with an immobile one, and $\rho_{im}$ represents the reactivation of the immobile dislocation due to stress or thermal activation. $a\rho_m$ represents the immobilization of mobile dislocations due to aggregation of solute atoms to dislocations. Once a mobile dislocation starts acquiring solute atoms we regard it as Cottrell-type dislocation $\rho_c$. As  more and more solute atoms aggregate, they eventually stop, and are considered as immobile dislocations $\rho_{im}$. This term then acts as a source term in Eq. (\ref{Z-eqn}). $\phi^m \rho_m $ represents the rate of multiplication of dislocations due to cross slip. This depends on the velocity of the mobile dislocations taken to be  $V_m(\phi) = \phi^m$, where $m$ is the velocity exponent.  Within the scope of the model, cross slip that allows dislocations to spread into neighboring spatial locations gives rise to diffusive coupling [the last term in Eq.  (\ref{X-eqn})]. These equations are coupled to  Eq. (\ref{phi-eqn}),  which represents the constant strain rate deformation  experiment. In Eq. (\ref{phi-eqn}), $\dot{\epsilon}_a$ is the scaled applied strain rate, $d$ the scaled effective elastic modulus of the machine and the sample, and $l$ the dimensionless length of the sample.  The scaled constants, $a,\, b_0$, and $c$ refer, respectively, to the concentration of solute atoms slowing down the mobile dislocation, the thermal and athermal reactivation of immobile dislocations, and the diffusion rate of solute atoms to the mobile dislocations. The drive  parameter is the applied strain rate $\dot{\epsilon}_a$ with respect to which the different types of bands and the associated serrations are observed. The instability range is found in the interval $20 < \dot{\epsilon}_a < 1625$ for a system size $N=100$.

Equations (\ref{X-eqn}$-$\ref{phi-eqn}) are discretized on a grid of $N$ points and solved using an adaptive step-size differential equation solver (MATLAB ''ode23'').  The initial values for the densities are taken to be  uniformly distributed  along the sample. However, as the long-term evolution does not depend on the initial values, steady-state values have been used. As dislocation bands can  not propagate into the grips, the boundary values $\rho_{im}(j,t); \, j=1$ and $N$ are  taken to be two orders higher than $\rho_{im}(j), j= 2,..N-1$. Further, we impose  $\rho_m(j,t) =\rho_c(j,t)=0$ for $j=1$  and $N$. The results reported are for $a=0.8, b= 5\times 10^{-4},  c=0.08, d= 6\times 10^{-5}, m=3,  D=0.25$. The system size  $N=20-150$ is used depending on the property addressed. After discarding the initial transients,  $\phi(t)$ is sampled at time intervals $\delta t=0.5$.

\section{Spatio-temporal Patterns and   Relation to Stress Serrations}

The results of the model relevant for the present purpose are the spatiotemporal patterns  reflected in the nature of  the dislocation bands and their connection to the changes in the nature of stress serrations, observed with increasing applied strain rate. To understand the origin of the instability and the nature of the spatial patterns, let us first examine Eqs. (\ref{X-eqn})  and (\ref{phi-eqn}) more carefully. We first note that  $\phi = \sigma/\sigma_y$, where $\sigma$ is the unscaled stress at time $t$ and $\sigma_y$ is the yield stress. From Eq. (\ref{X-eqn}), it is clear that   $\rho_m$ increases abruptly only when $\phi$ exceeds unity (corresponding to the unscaled stress exceeding the yield stress) or equivalently, there is a threshold for nucleation of an isolated  burst of the mobile density $\rho_m$.  However, a stress drop can only  occur when the space-averaged plastic strain rate $\dot \epsilon_p=\frac{\phi^m(t)}{l}\int_0^l \rho_m(x,t) dx$ exceeds the applied strain rate $\dot\epsilon_a$.

\begin{figure}[t]
\vbox{
\includegraphics[height=4.0cm,width=8.0cm]{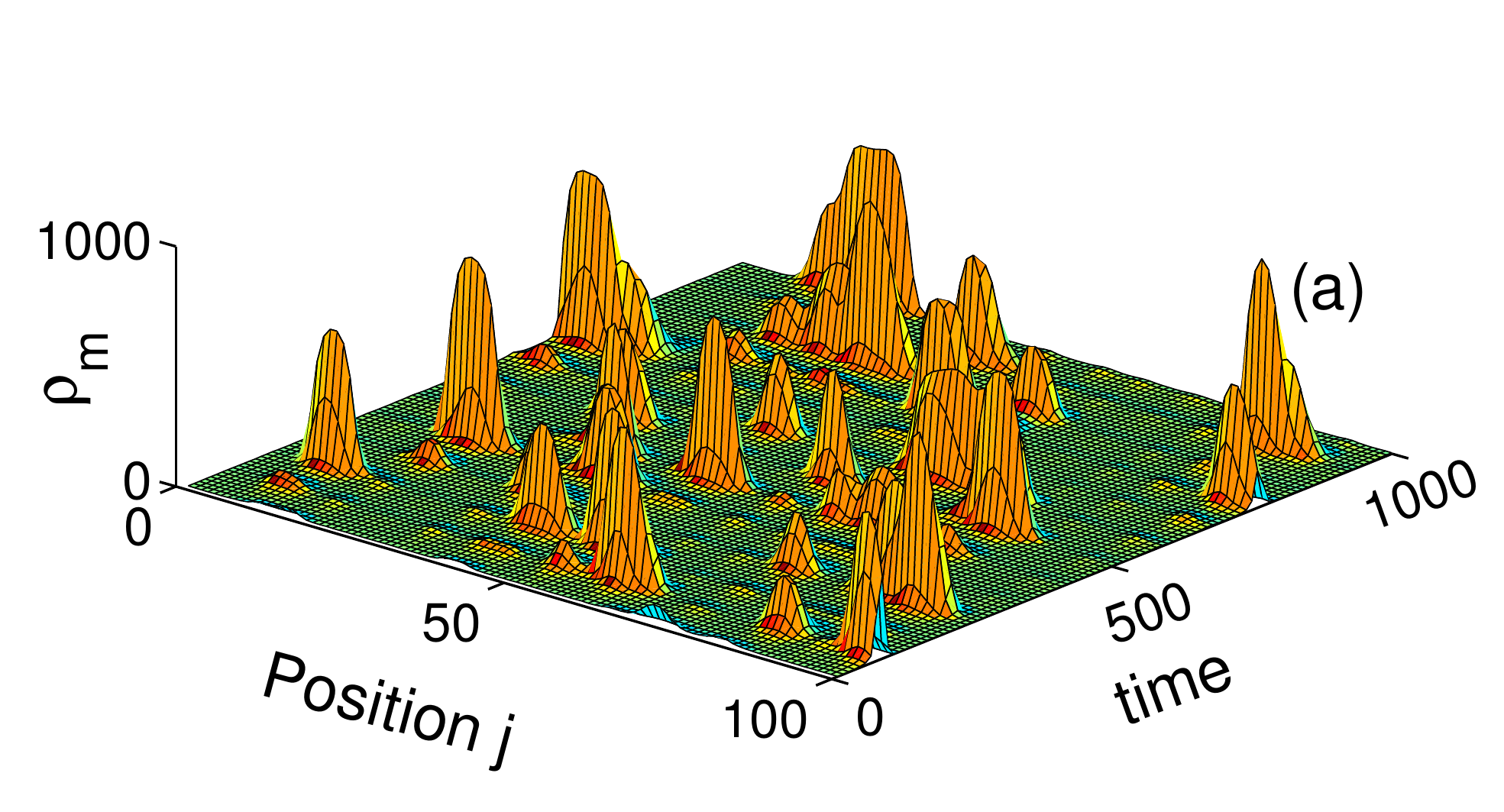}
\includegraphics[height=3.3cm,width=7.5cm]{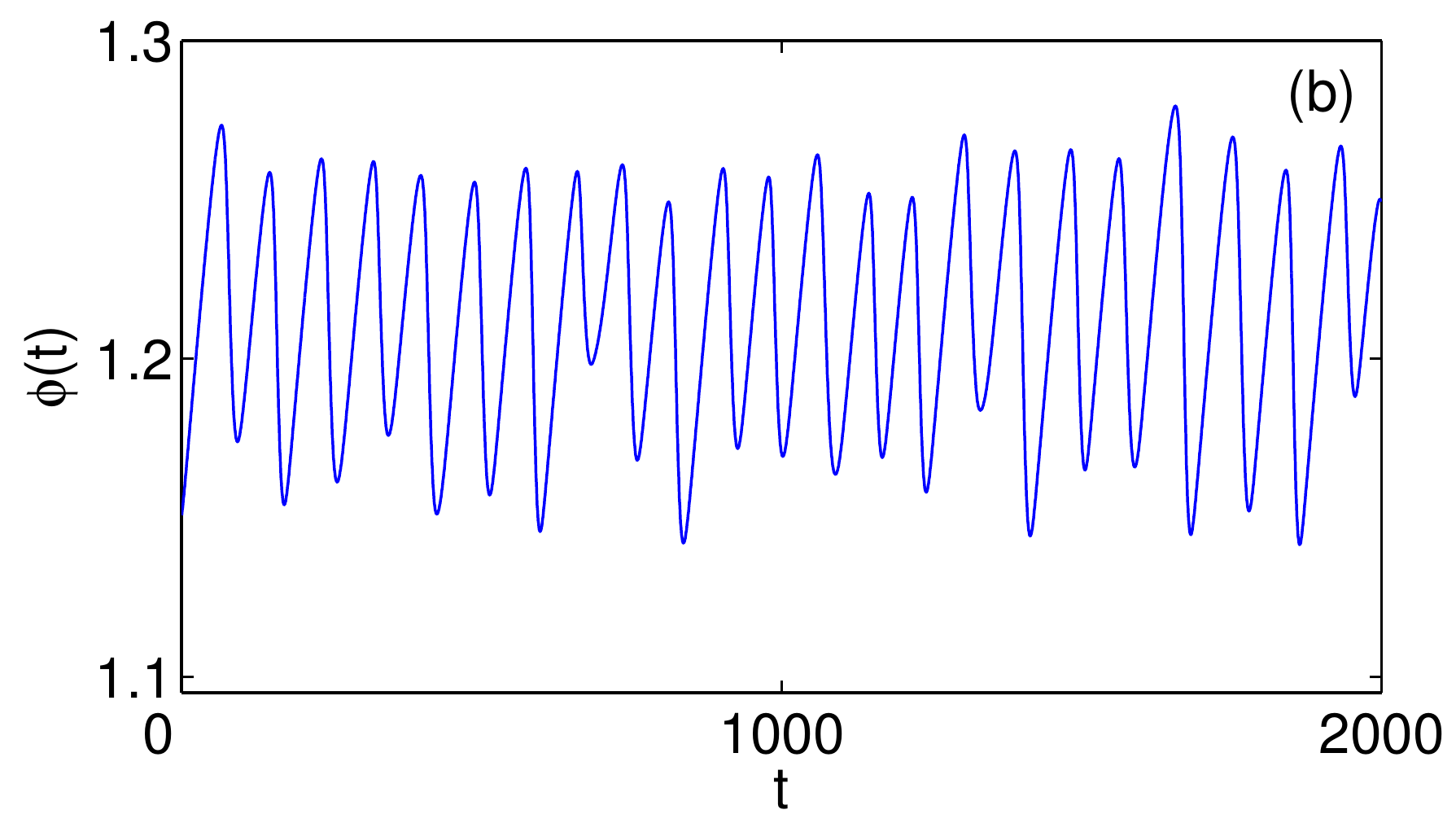}
}
\caption{(Color online) (a) Randomly nucleated type-C bands for $\dot\epsilon_a = 40$. (b) The corresponding stress-time plot. 
}
\label{Band40}
\end{figure}

\begin{figure}[h]
\vbox{
\includegraphics[height=4.0cm,width=8.0cm]{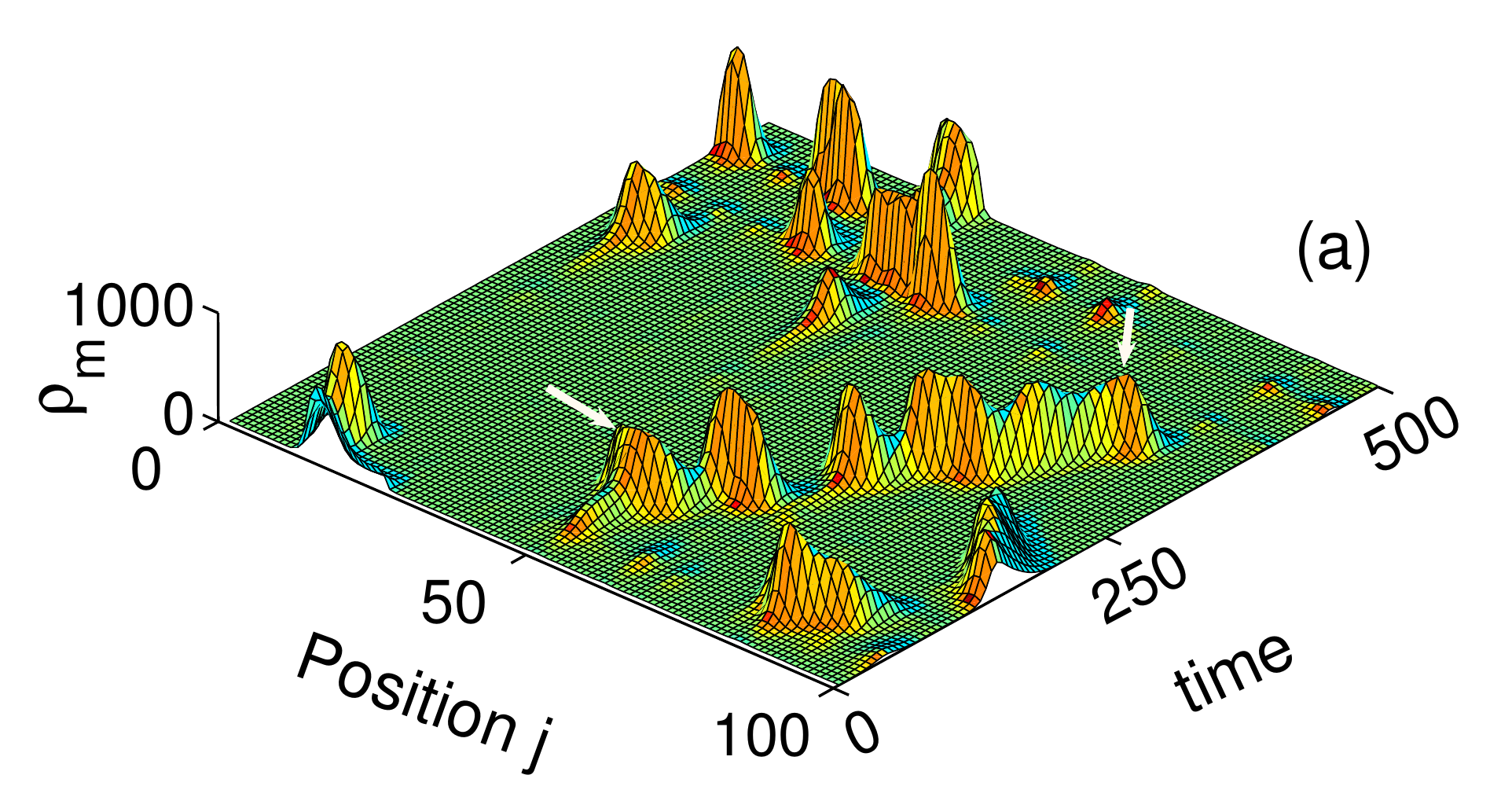}
\includegraphics[height=3.3cm,width=7.5cm]{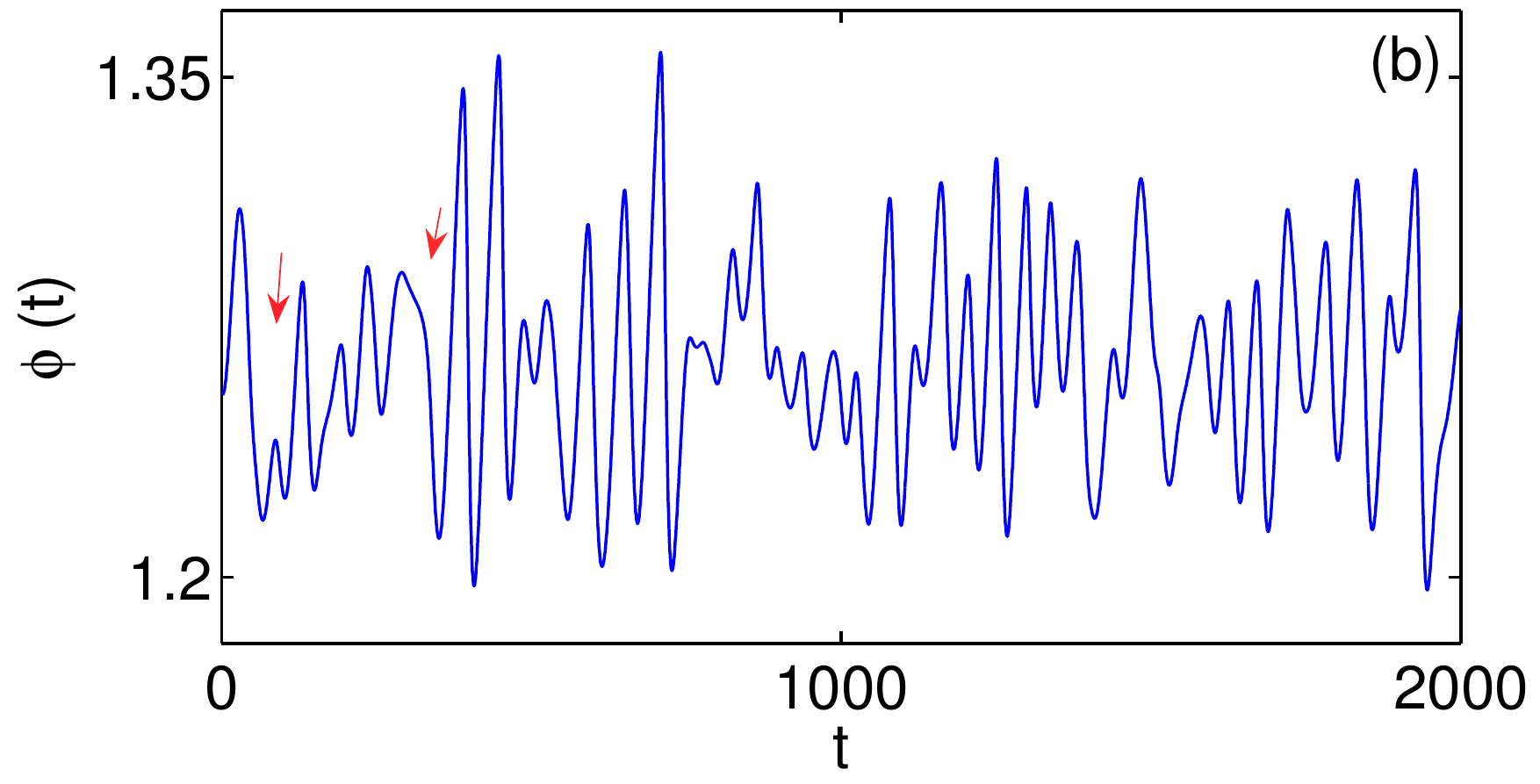}
}
\caption{(Color online) (a) Hopping  type-B bands for $\dot\epsilon_a = 90$. (b) The corresponding stress-time plot. For the sake of clarity, a short segment of the space-time plot is displayed. The correspondence between the propagating nature of the band marked by the  arrows in (a) and the stress-time series in (b) is displayed.  }
\label{Band90}
\end{figure}

\begin{figure}[h]
\vbox{
\includegraphics[height=4.0cm,width=8.0cm]{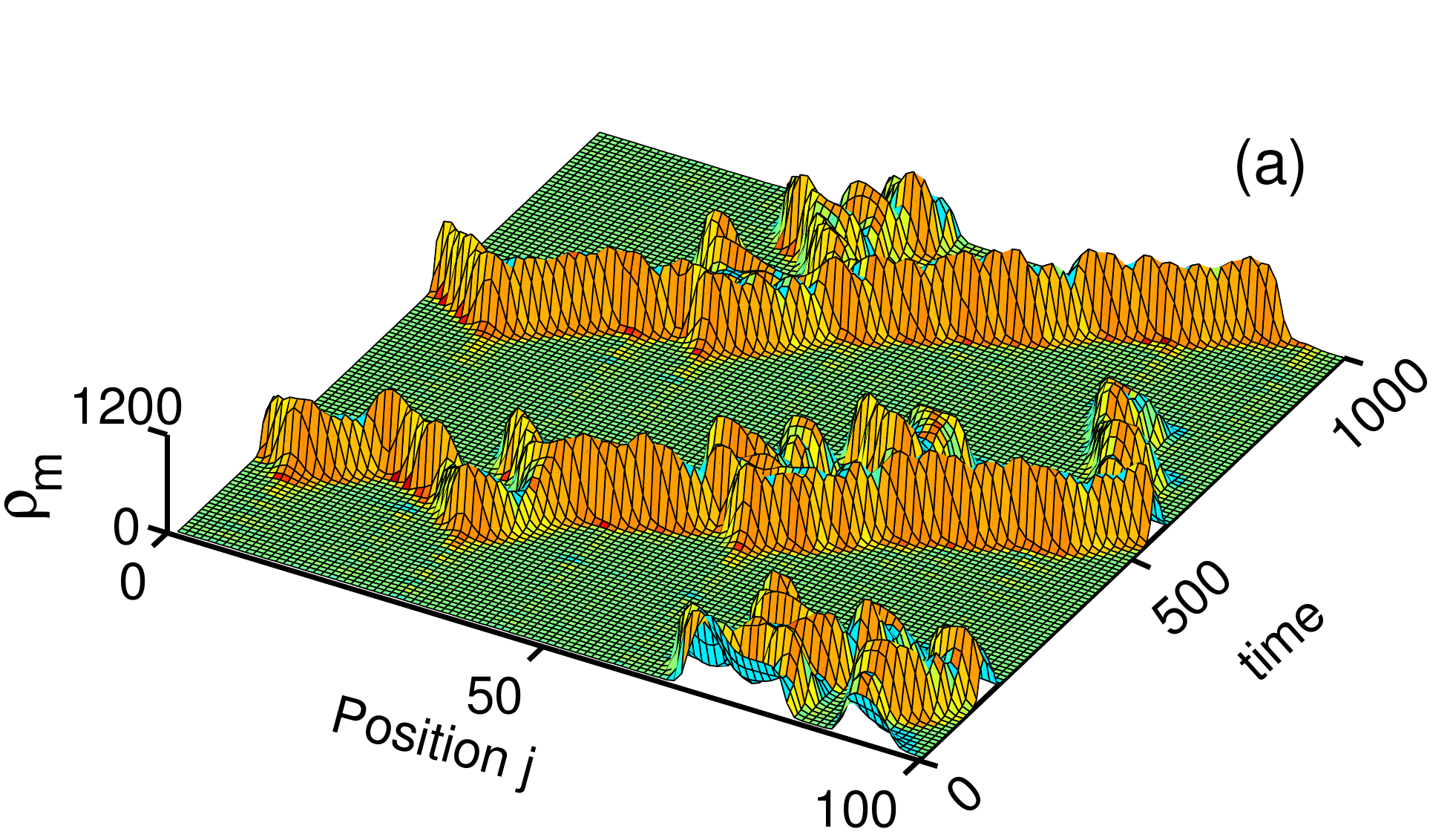}
\includegraphics[height=3.3cm,width=7.5cm]{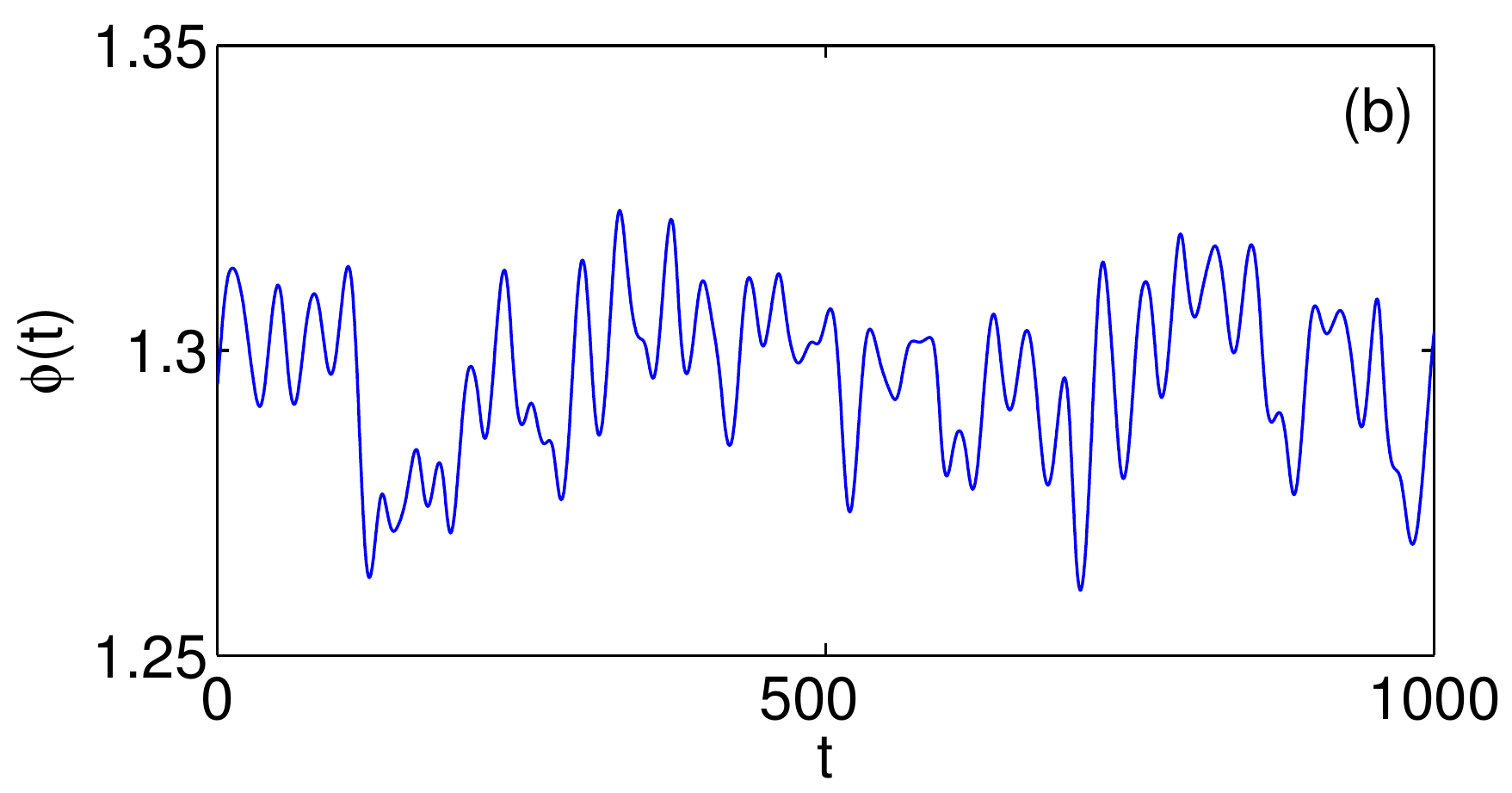}
}
\caption{(Color online) (a) Continuously propagating bands for $\dot\epsilon_a = 240$. (b) The corresponding stress-time plot.  }
\label{Band240}
\end{figure}

As mentioned earlier the model reproduces all the three types of bands (C, B, and A) observed in experiments with increasing $\dot\epsilon_a$ \cite{GA07,Bhar03b}. For a range of low $\dot\epsilon_a$ values from 20 to 50,  randomly nucleated dislocation bands are seen in the form of isolated bursts of $\rho_m$. From the above discussion,  we know that an isolated   burst of $\rho_m$ can only be nucleated if the scaled stress $\phi$ increases beyond unity.  A typical space-time plot of the randomly nucleated bands is shown in Fig. \ref{Band40}(a) for $\dot\epsilon_a=40$.  The corresponding stress-time curve is nearly regular as illustrated in Fig. \ref{Band40}(b).  Furthermore, in this range of $\dot\epsilon_a$, usually a single burst of $\rho_m$ contributes to the total plastic strain rate $\dot \epsilon_p=\frac{\phi^m(t)}{l}\int_0^l \rho_m(x,t) dx$  as the magnitude of $\rho_m(x,t)$ at other spatial locations   is insignificant.  This implies that whenever the space-averaged plastic strain rate overshoots  the applied strain rate $\dot\epsilon_a$, a stress drop occurs [see Eq. (\ref{phi-eqn})]. Thus, there is a one-to-one correspondence between the burst of mobile density $\rho_m$  and the stress drop. However, as we increase $\dot\epsilon_a$, a new burst of $\rho_m$ is formed even before the previous burst dies off giving the impression of hopping character (type-B band). This is clear from the  plot of the hopping-type partial propagating  band  shown in Fig. \ref{Band90}(a) for $\dot\epsilon_a=90$.  This also means that the stress required to nucleate a fresh burst of $\rho_m$ ahead of the previous one even before it dies off, is less than that required to nucleate an isolated burst of $\rho_m$. This in turn implies that the amplitude of the serrations would be smaller in regions where propagating bands are seen.  This relationship between the propagative nature of the bands and small-amplitude stress drops is illustrated in Figs. \ref{Band90}(a) and \ref{Band90}(b). The figure  identifies the propagative regime with the corresponding stretch of  small amplitude stress serrations [marked by a corresponding set of arrows in the  Fig. \ref{Band90}(a) and \ref{Band90}(b)]. As we increase $\dot\epsilon_a$, the extent of propagation increases, concomitantly, longer stretches of small-amplitude serrations interrupt the otherwise large-amplitude serrations.  The spatial correlation increases with $\dot{\epsilon}_a$ until the bands propagate fully with numerous small stress drops. The corresponding stress-time plot would have mostly small-amplitude stress drops with very few large stress drops. A typical plot of a fully propagating band  along with the associated stress-time plot is shown in Figs. \ref{Band240}(a) and \ref{Band240}(b). (see Refs. \cite{GA07,Bhar03b}).

\section{Spatio-temporal Dynamics of the Model}

\begin{figure}[t]
\vbox{
\includegraphics[height=4.5cm,width=8.0cm]{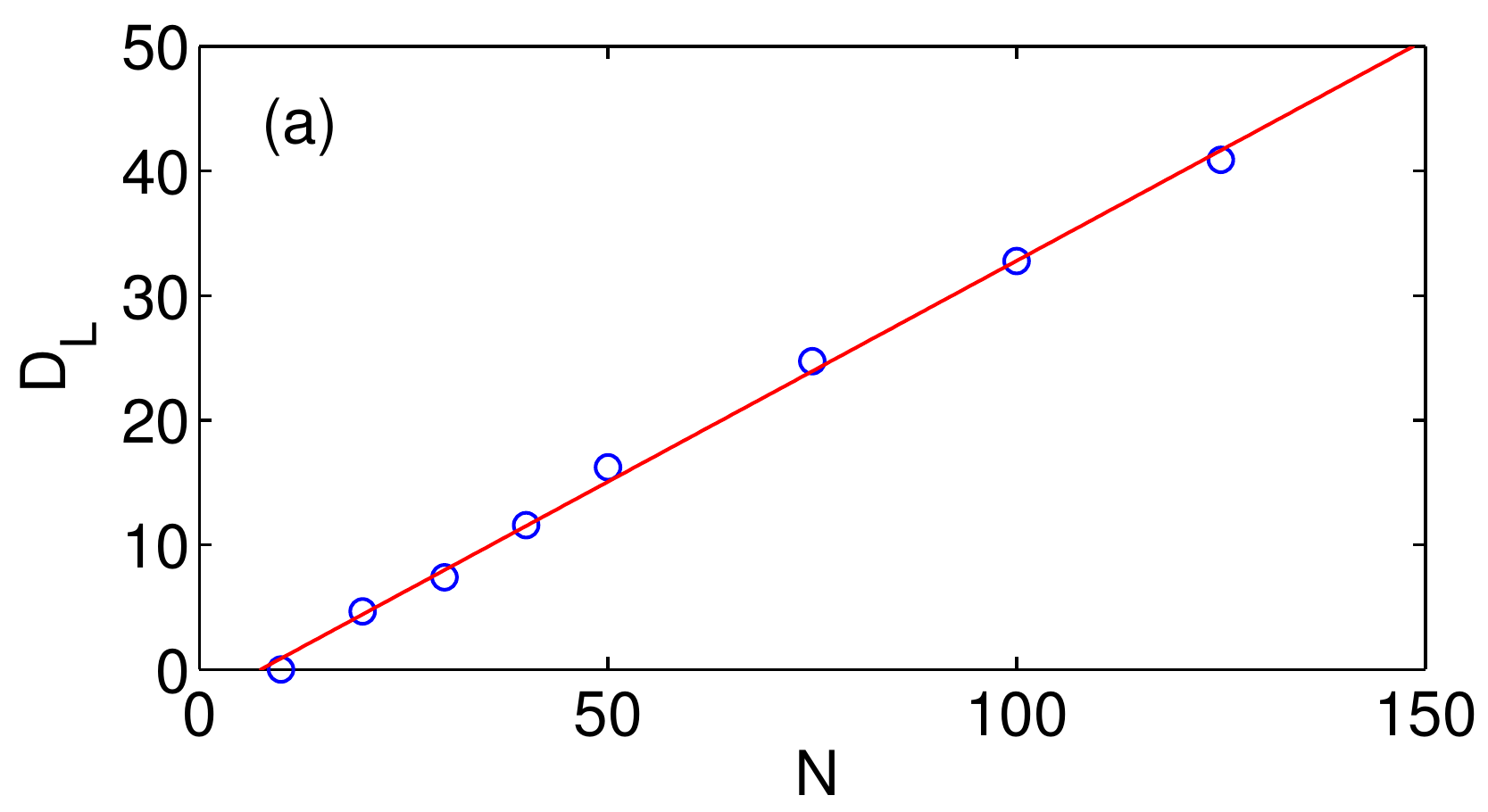}
\includegraphics[height=4.5cm,width=8.0cm]{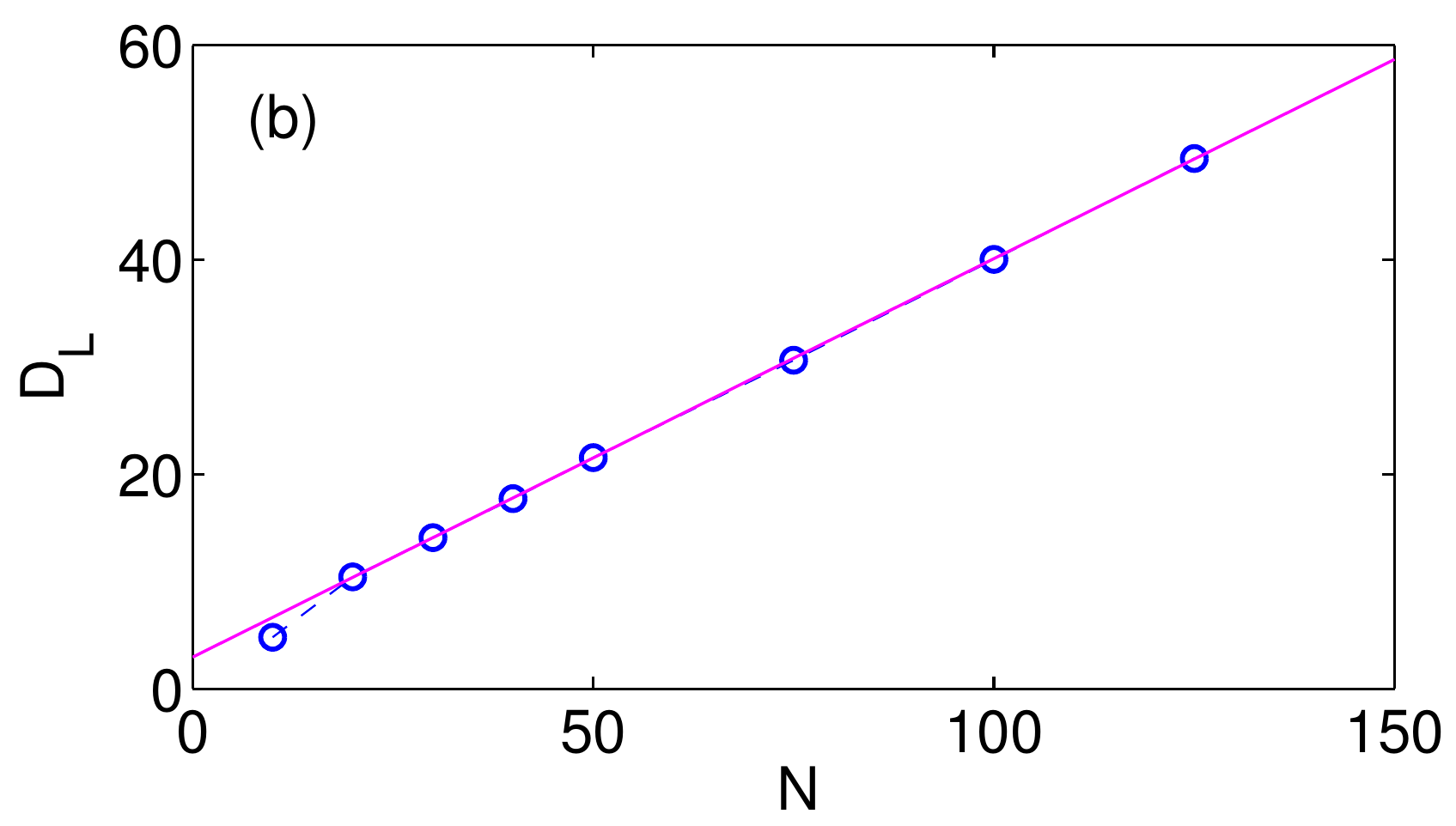}
}
\caption{(Color online) Lyapunov dimension $D_L$ as a function of $N$ (a) for $\dot\epsilon_a=40$ (randomly nucleating bands) with slope $\approx 0.35$ and (b) for $\dot\epsilon_a=240$ (continuously propagating bands) with slope $\approx 0.37$. }
\label{DL-N-40-240}
\end{figure}

\begin{figure}[h]
\vbox{
\includegraphics[height=4.5cm,width=8.0cm]{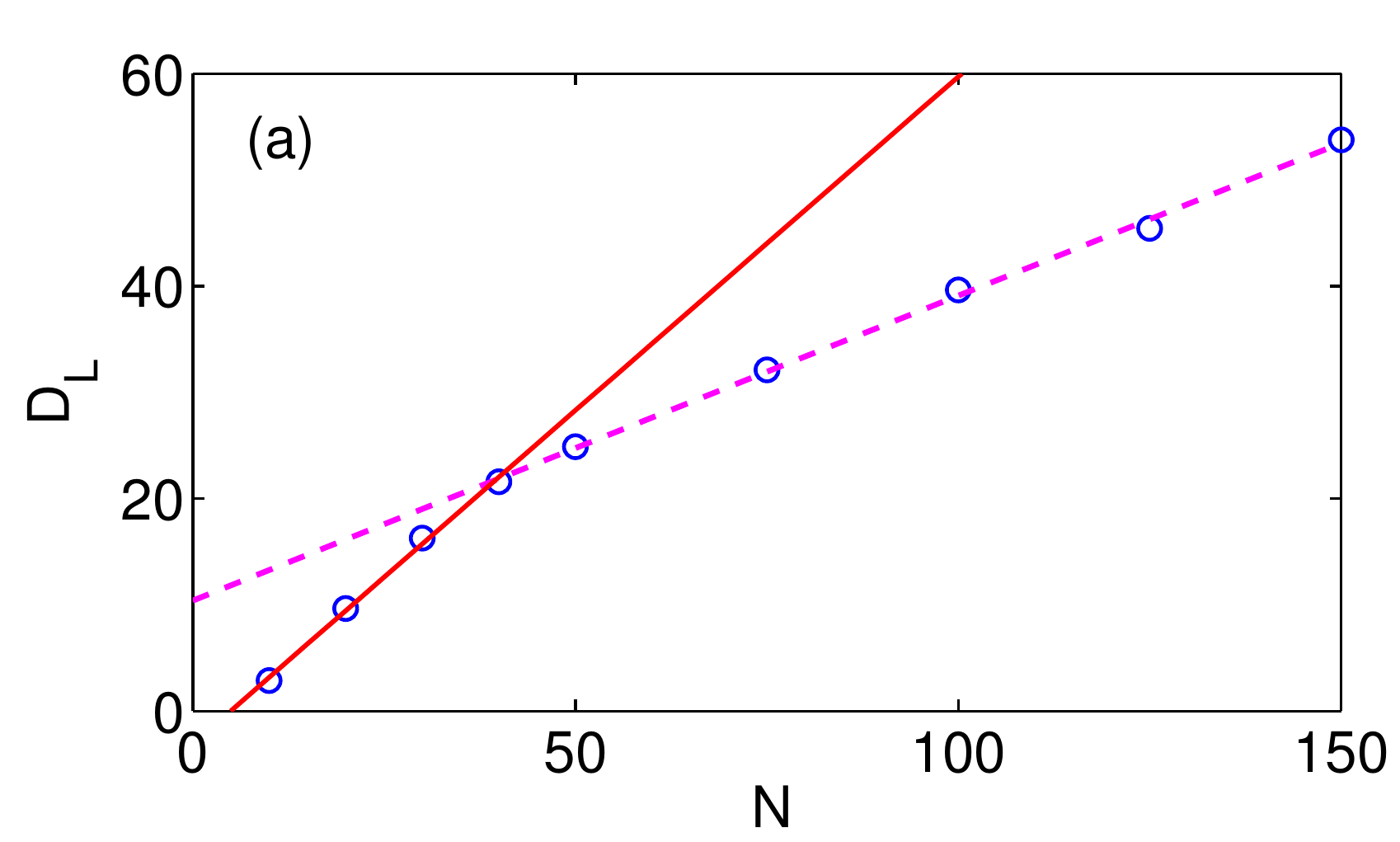}
\includegraphics[height=5.0cm,width=8.0cm]{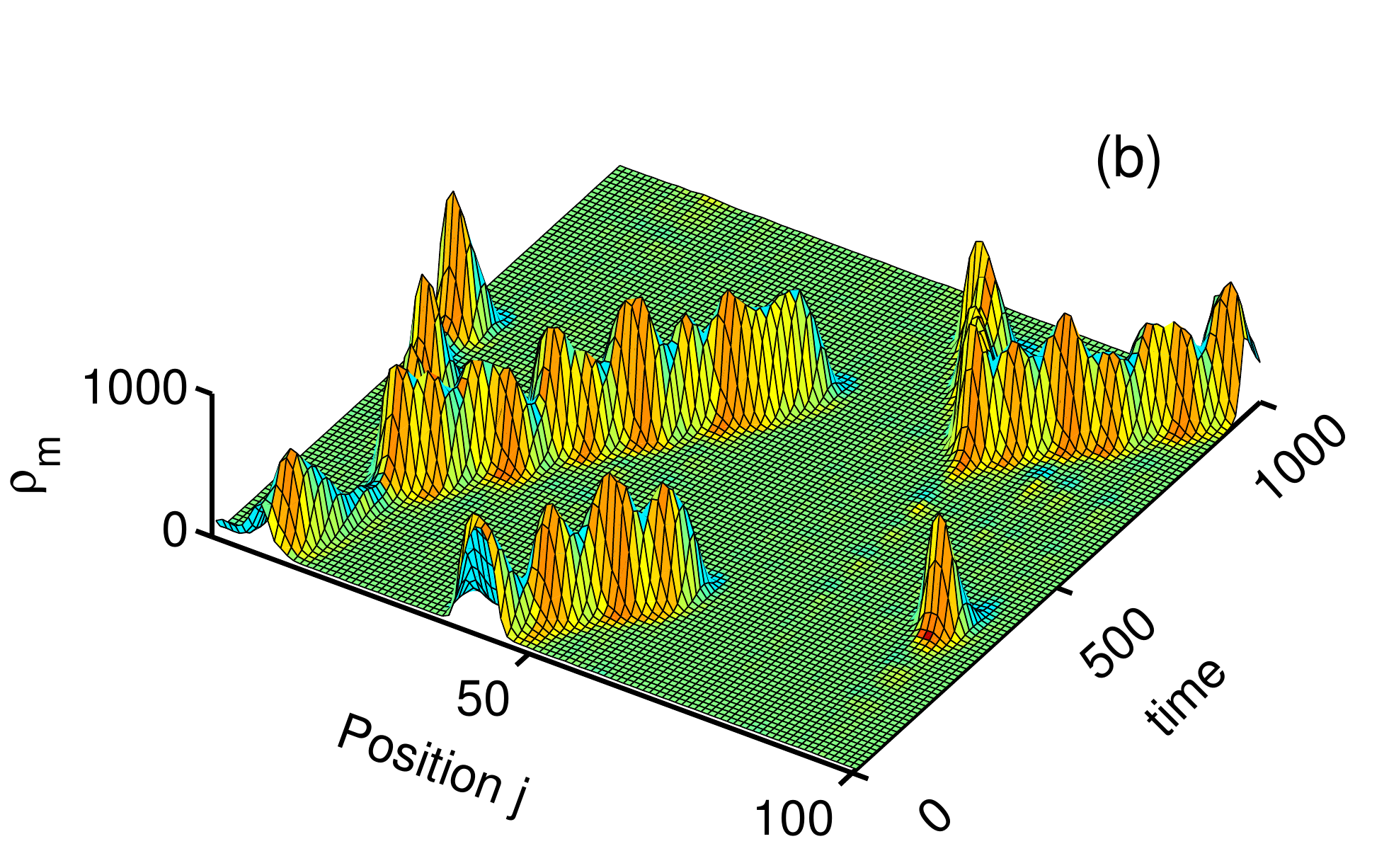}
\includegraphics[height=5.0cm,width=8.0cm]{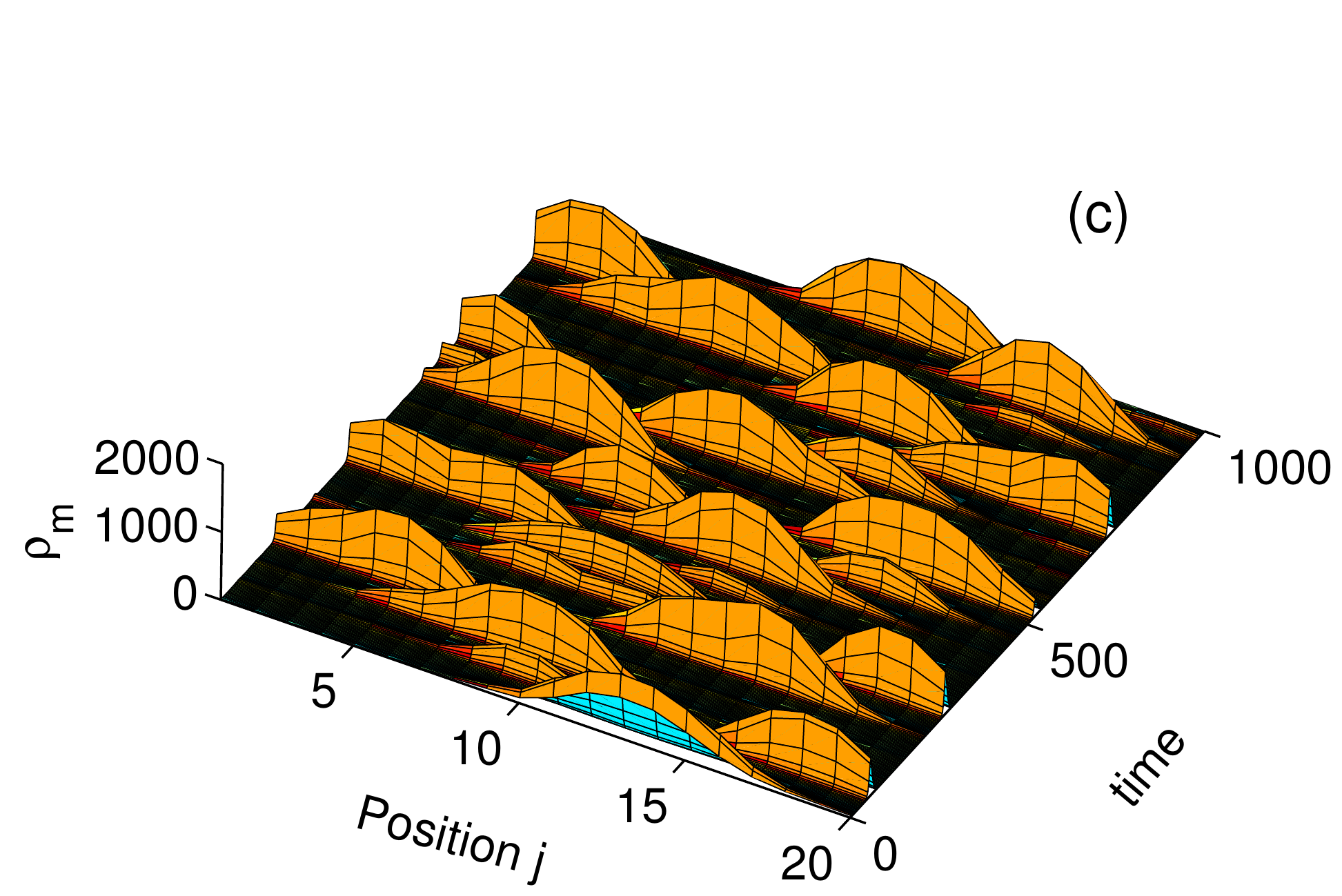}
}
\caption{(Color online) (a) Lyapunov dimension $D_L$ as a function of $N$ for $\dot\epsilon_a=120$. Note the two distinct slopes. One for larger system sizes ($N>50$) with slope $\approx 0.287$ (dashed line) and the other for  smaller system sizes with slope $\approx 0.67$ (continuous line) (b) Space-time plot of the mobile dislocation density for the same $\dot\epsilon_a$ and $N=100$. (c) Space-time plot of the mobile dislocation density for  $\dot\epsilon_a =120$ and $N=20$.}
\label{DL-N-SR120}
\end{figure}

The above spatio-temporal dynamics has been quantified by calculating the Lyapunov spectrum from the model equations (using the Bennettin's algorithm \cite{Ben80}) for a range of values of $\dot\epsilon_a$ where the randomly nucleated bands and partial propagative bands are seen. (First $10,000$ points have been ignored and the spectrum is calculated using the next $10,000$ time steps.)   We have studied the system size dependence of the number of  positive exponents $n_{\lambda}^+$ and the 
Lyapunov dimension $D_{L}$. Both scale linearly with $N$ for low strain rates where randomly nucleating bands are seen and also at high strain rates where fully propagating bands are seen. For illustration, plots of $D_L$ verses $N$ is shown in Figs. \ref{DL-N-40-240}(a) and \ref{DL-N-40-240}(b), for $\dot\epsilon_a=40$ and $\dot\epsilon_a=240$, respectively. However, for a range of values of $\dot\epsilon_a$ beyond $60$ where partial propagative bands are seen [that also coincides with the region where we do not find converged values of the correlation dimension, $D_{2s}(r,d)$, as we shall soon show], we find two distinct slopes in the $D_L-N$ plot, one for small values of $N$ and another for large values of $N$. This feature can be seen even for $\dot\epsilon_a=60$ where partial propagation begins. A typical plot of $D_L-N$ is shown in Fig. \ref{DL-N-SR120}(a)  for $\dot\epsilon_a=120$. (The larger slope for small $N$ values is $\sim 0.67$ and the smaller slope for large $N$ is $\sim 0.287$.) As we increase the strain rate, the range of values of $N$ corresponding to the larger slope decreases; eventually a single slope is seen for $\dot\epsilon_a=240$ [Fig. \ref{DL-N-40-240}b]. A similar behavior is also seen for $n_{\lambda}^+$. 

To understand the underlying reasons for the two slope nature of $D_L$ versus $N$, we  have examined  the nature of the spatiotemporal patterns when the system size is increased from small values of $N$ to large values. Plots of spatiotemporal patterns for $\dot\epsilon_a=120$ for $N=100$ and $N=20$, are shown in Fig. \ref{DL-N-SR120}(b) and \ref{DL-N-SR120}(c), respectively. As can be seen from the two plots, while  the bands propagate fair distances for large system size $N=100$, for small system size $N=20$, one finds well-separated  bursts of $\rho_m$ (in the time domain) at any given time, although there is a visual impression of propagation. However, the direction of apparent propagation [in Fig. \ref{DL-N-SR120}c] is opposite to that for $N=100$ [Fig. \ref{DL-N-SR120}b]. Indeed, we find that the dynamics is altered from partial propagative nature for large system size (say for $N=100$) to one of burst type when the system size is smaller than $N=50$. (Note that the burst-type pattern is seen  for  $\dot\epsilon_a < 60$ when $N$ is large.) Similar altered dynamics for small system sizes is seen for $ 60 \le \dot\epsilon_a \le 200$.

\section{Time Series Analysis of Stress Signals}

Now consider analyzing the stress-time series that is related to the spatial average of the dislocation activity in the sample [Eq. \ref{phi-eqn}]. Consider a stress-time series of length $M$ in units of $\delta t$ defined by $\{\phi(k), k=1,2,\cdots,M \}$. Then, the reconstructed attractor is defined by the $d_E$ dimensional vectors $\vec{\xi}_k = \{\phi(k), \phi(k+\tau),\cdots,\phi[k+(d_E-1)\tau]\}$; $k=1, 2,\cdots,[M-(d_E-1)\tau$], where $\tau$ is the delay time. The chaotic nature of the attractor is quantified by establishing the existence of finite correlation dimension and a positive Lyapunov exponent.

The most popular method for calculating the correlation dimension is the Grassberger-Procaccia (GP) algorithm \cite{GP83}. The method calculates the correlation integral defined as the fraction of the pairs of points $\vec{\xi}_i$ and $\vec{\xi}_j$ whose distance is less than a specified value $r$, that is, 
\begin{equation}
C(r) = \frac{1}{M_p} \sum_{i,j} \Theta(r- \vert \vec{\xi}_i-\vec{\xi}_j\vert ),
\end{equation} 
where $ \Theta(...)$ is the Heaviside step function and $M_p$ the number of vector pairs used \cite{GP83}. A window is imposed to exclude temporarily correlated points. The correlation dimension is  defined as
\begin{equation}
D_2(r,d_E)= \lim_{r \rightarrow 0} \lim_{M \rightarrow \infty} \frac{ d ln \, C_2(r,d_E)}{dln \,r}.
\end{equation}
This limit is seldom reached. Instead, one usually finds a reasonably large scaling regime in $\ln \, r$ at intermediate scales, where the slope $D_2(r,d_E)$ converges to a finite value $D_2$, which is taken to be the correlation dimension of the attractor.  The method has been successfully applied for analyzing  several (mostly) low-d attractors \cite{Kantz97}. However, the validity of the GP algorithm for high dimensional attractors has been questioned since  exponentially longer time series are required \cite{Smith88, Ruelle92, Dvorak90}. Generally, two types of errors limit the confidence in the estimates of $D_2$, namely the poor statistics at small length scales and the underestimation of $D_{2}(r,d_E)$ at length scales comparable to the attractor size. The latter arises when the reference point is close to the edge of the attractor as zero contribution arises for length scales $r$  beyond the attractor's edge. This systematic error is shown to be important in  characterizing  high dimensional attractors including those of  spatially extended systems \cite{Bauer93, Kurths01}.
Bauer {\it et al.} \cite{Bauer93} suggested a method of compensating the contribution arising from finite size of the attractor by normalizing the local slope $D_{2}(r,d_E)$ by the slope of an equivalent random attractor $D_{2r}(r,d_E)= \frac{d \,ln \, C_{2r}(r,d_E)}{d \, ln\, r}$. Thus, the ``true`` correlation dimension is redefined as,
\begin{equation}
 D_{2s}(r,d_E) = \frac{D_2 (r, d_E)}{D_{2r}(r, d_E)} = \frac{d \,ln \,C_2(r,d_E)}{d_E \, d \, ln C_{2r}(r,1)} 
\label{ModD2}
\end{equation}
where $C_{2r}(r,d_E) = C_{2r}(r,1)^{d_E}$  has been used\cite{Bauer93}. Here we use $C_2(r,2)^{1/2}$ instead  of $C_{2r}(r,1)$ and define 
\begin{equation}
 D_{2s}(r,d_E)= \frac{D_2(r,d_E)}{D_{2}(r,2)/2} = 2\frac{d \, ln \, C_2(r,d_E)}{d \, ln C_2(r,2)}.
\end{equation} 
Using $C_2(r,2)^{1/2}$ not only serves to correct for the finite size of the attractor as does $C_{2r}(r,1)$, it also includes the contribution from finite delay time. 
The converged value of $D_{2s}(r,d_E)$ over a fair range of $ln \, r$ is taken to be $D_2$. We further use constant window length $(d_E-1)\tau$ that maximizes the scaling regime \cite{Albano88}. The Lyapunov spectrum for the time series is computed using the standard Eckmann's algorithm \cite{Eckmann86}.

\begin{figure}
\includegraphics[height=4.5cm,width=8.0cm]{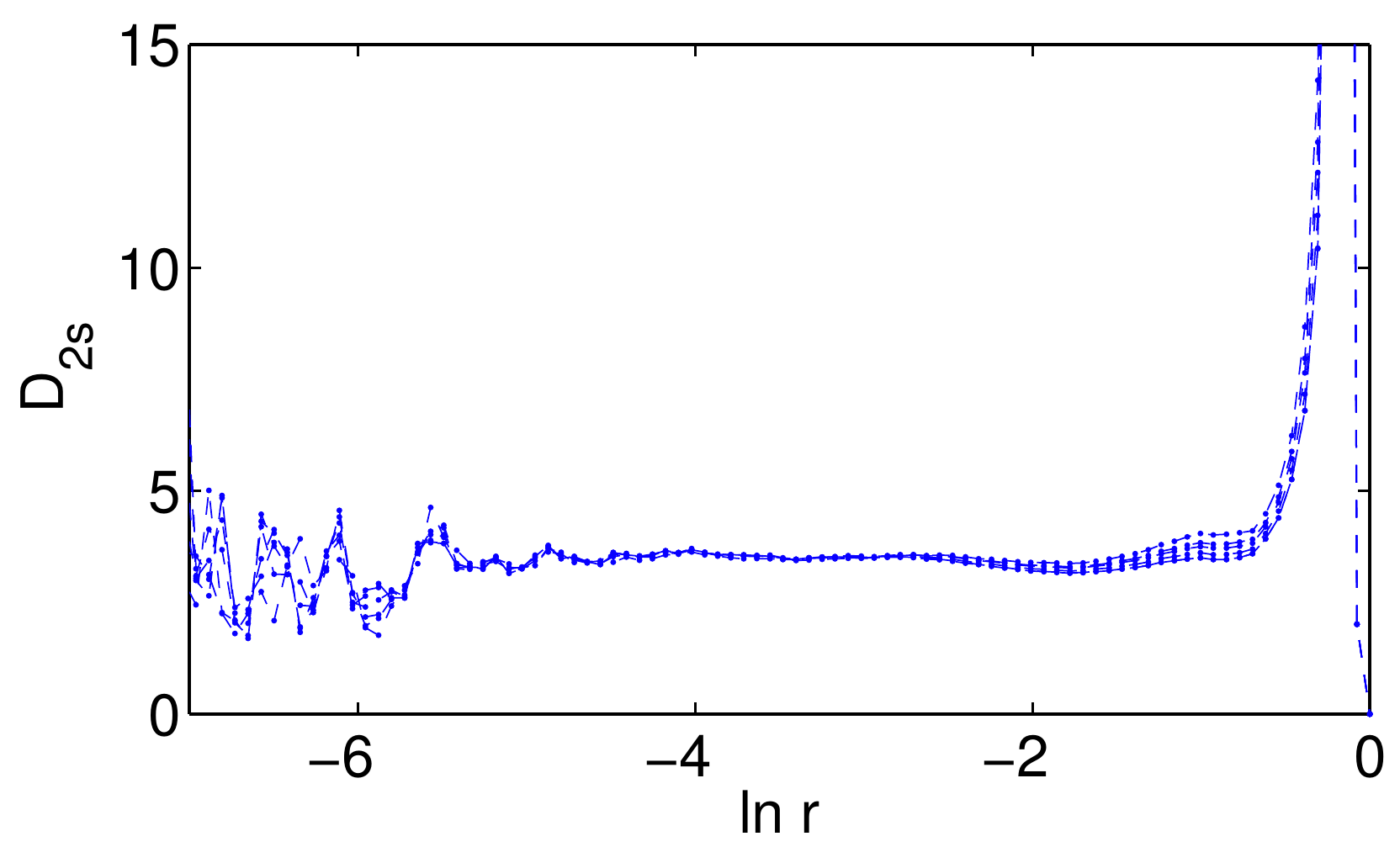}
\caption{ (Color online)  Plot of $D_{2s}(d_E,r)$ as a function of  $ ln\,r$ for the sum of two independent $x$ components of the Lorenz model for $(d_E,\tau) = (5,40),(6,32),(7,27),(8,23),(9,20),(10,18)$. Normalization used is $D_2(r,2)/2$ for $\tau=250$.}
\label{TLorentz}
\end{figure}

\begin{figure}
\vbox{
\includegraphics[height=4.5cm,width=8.0cm]{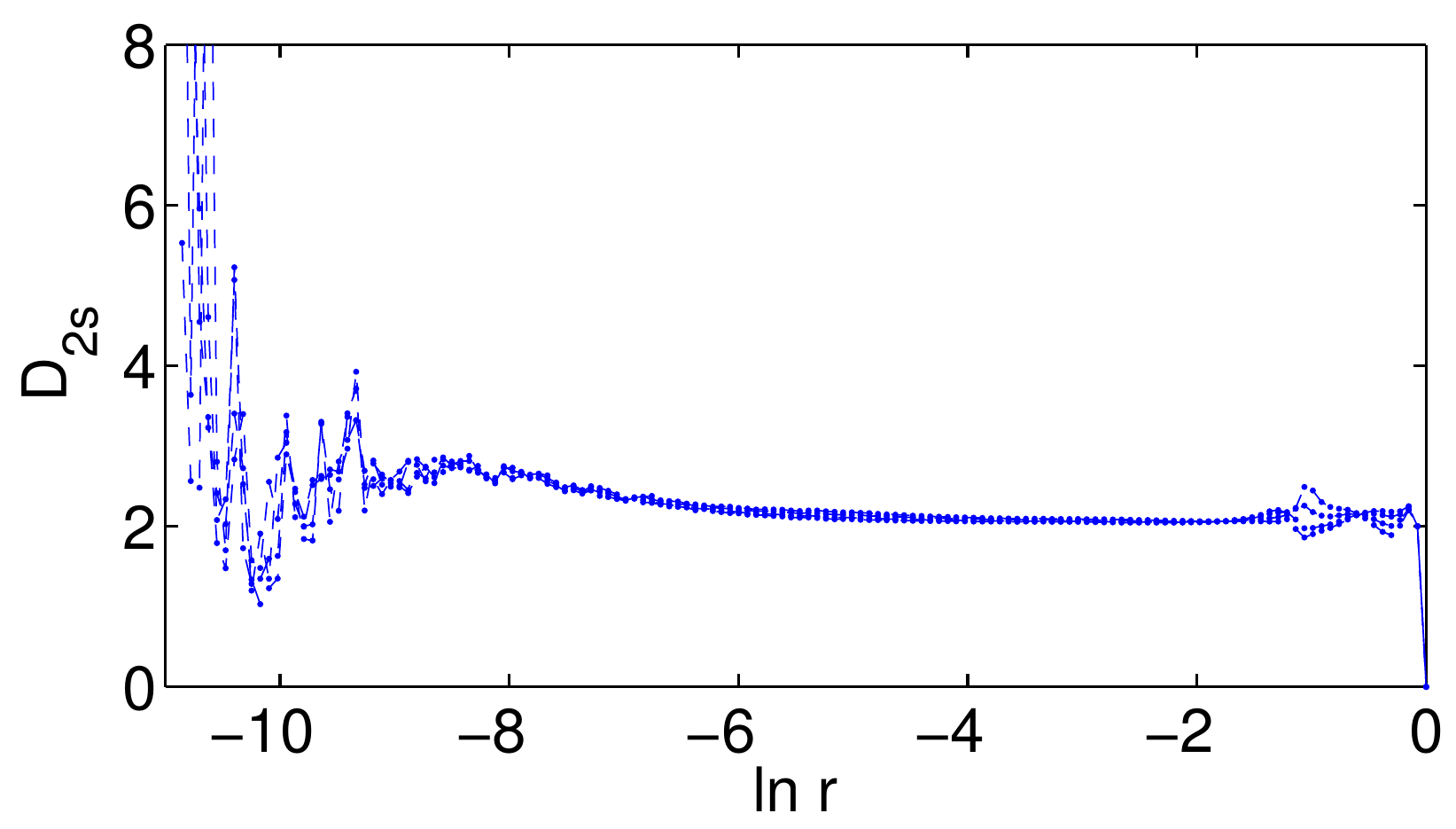}
}
\caption{(Color online)  Plot of $D_{2s}(r, d_E)$ as a function of  $ ln \, r$ for the stress-strain series for $\dot\epsilon_a=25$ for $1.1\times 10^5$ points. The values of $(d_E,\tau)$ are (4,16),(5,12),(7,8),(9,6). Normalization used is $D_2(r,2)/2$ for $\tau=30$.}
\label{Corrd25}
\end{figure}

We first check the  applicability of the algorithm  to high dimensional attractors with dimensions, say four or five. For this, we consider  a time series obtained by  summing two independent  x-component time series of the Lorenz model. The first data set is for parameter values $\sigma=10,r=28,b=8/3$ and the second is for $\sigma=16,r=40,b=8/3$. The total number of points used is $10^5$. We have kept  $(d_E-1) \tau =160$, an optimum value of the window length that  maximizes the  scaling regime to calculate the $C_2(r,d_E)$ for $(d_E,\tau) = (5,40),(6,32),(7,27),(8,23),(9,20),(10,18)$.   The slope $D_2(r,d_E)$ has been normalized with respect to $D_2(r,2)/2$ for $\tau=65$.  Figure \ref{TLorentz} shows the plot of $D_{2s}(r,d_E)$ for this case. It is clear that there is nearly five orders of scaling regime, which is comparable to the scaling regime obtained using an adoptive box counting method where $10^7$ points have been used \cite{Corana}. The value of $D_2$ obtained is $\sim 3.8$. Note that $ 2 D_2 = 4.12$ is only the upper limit.  The above algorithm works very well for low-d attractors such as the Lorenz model (where we find six orders of scaling regime 
with just $10,000$ points).

\begin{figure}
\vbox{
\includegraphics[height=4.5cm,width=8.0cm]{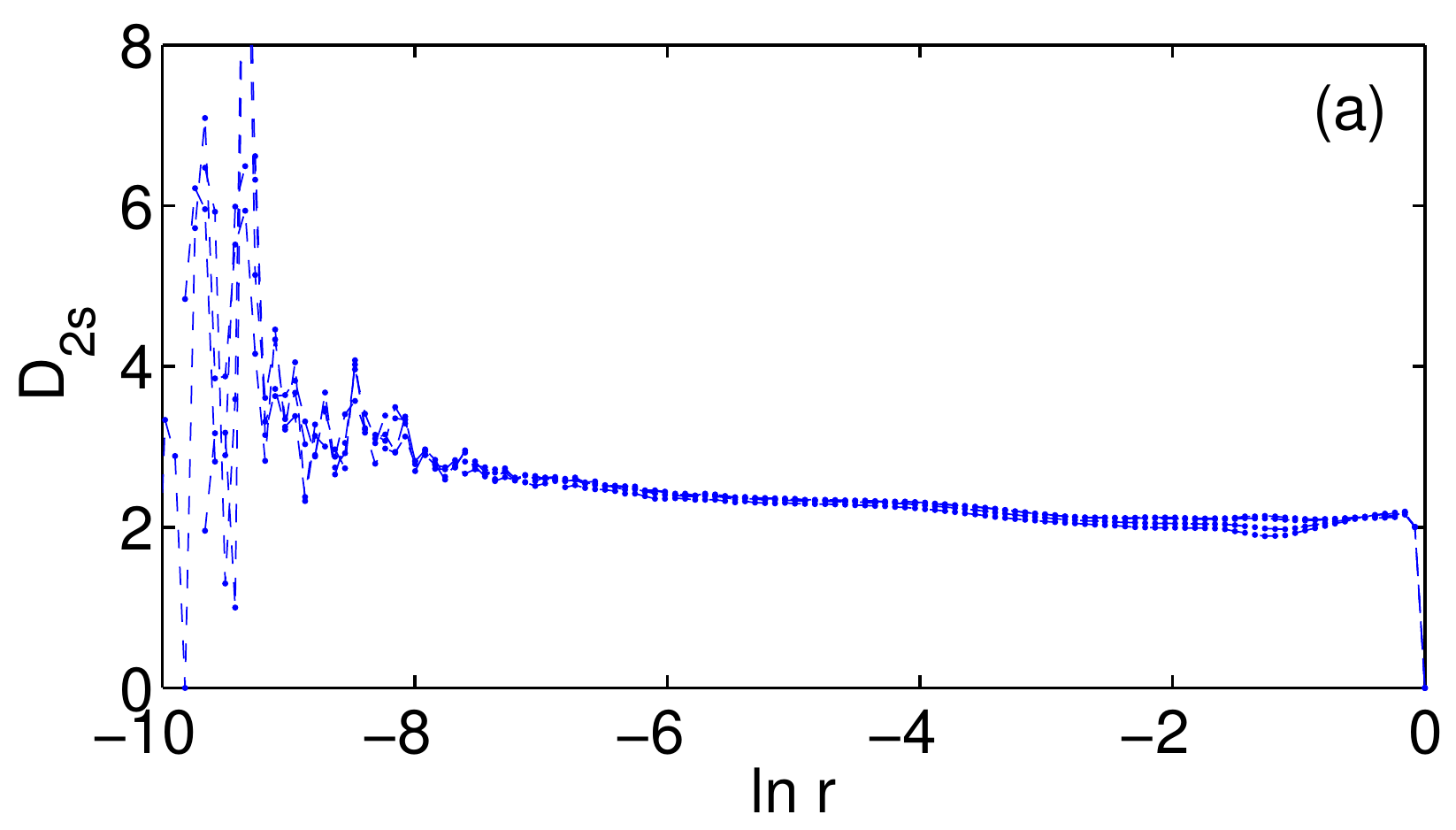}
\includegraphics[height=4.5cm,width=8.0cm]{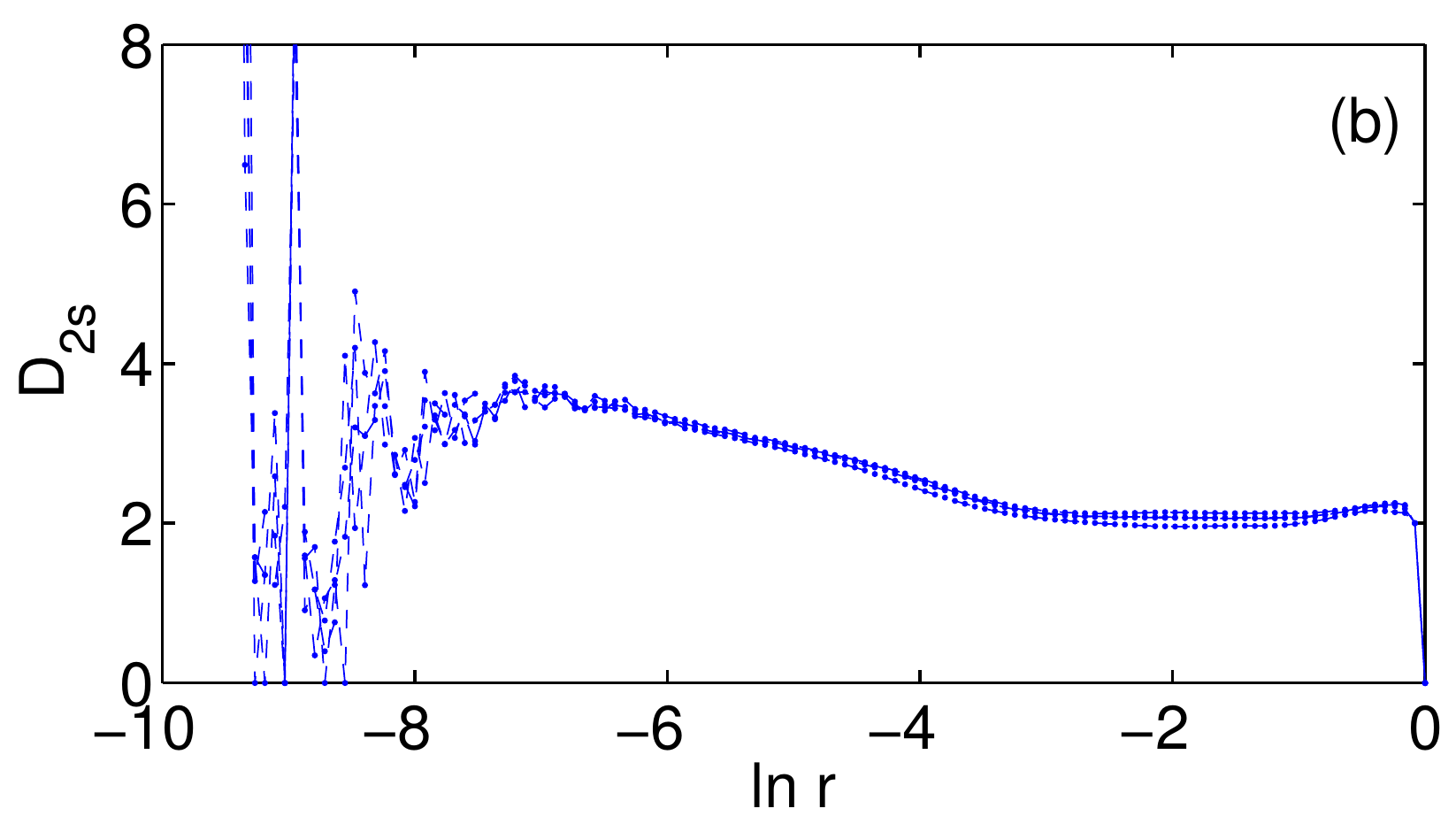}
}
\caption{(Color online)(a)  Plot of $D_{2s}$ versus $ ln \, r$ for the stress-time series for $\dot\epsilon_a=60$. The values of $(d_E, \tau)$ are such that $(d_E -1)\tau = 27$. Normalization used is $D_2(r,2)/2$ for $\tau=18$. (b) Plot of $D_{2s}$ versus $ ln \, r$ for the stress-time series for $\dot\epsilon_a=90$. The values of $(d_E, \tau)$ are such that $(d_E -1)\tau = 24$. Normalization used is $D_2(r,2)/2$ for $\tau=16$.}
\label{Corrd90}
\end{figure}

Having demonstrated the  spatiotemporally chaotic nature of the model equations, we now examine if the stress-time series has the required features of low-dimensional chaos using the above algorithm. Here we note that the stress rate $\dot \phi$ depends linearly on the  plastic strain rate which  in turn involves the spatial average of mobile dislocation density. Thus, the scalar signal $\phi$, though a dynamical variable, which determines the rate of multiplication of mobile density [first term in Eq. (\ref{X-eqn})], is an appropriate quantity for carrying out time series analysis. We have  calculated $D_2$ from the  stress-time series $\phi(t)$ in the range $ 20<\dot\epsilon_a \le 50$ corresponding to the randomly nucleating bands using $1.1 \times 10^5$ points for $N=100$.  A plot of $D_{2s}(r,d_E) $  as a function of $ ln \, r$  for $\dot\epsilon_a = 25$ is  shown in Fig. \ref{Corrd25}, keeping $(d_E-1)\tau=48$ ($d_E = 4,5,7$ and $9$) \cite{Albano88}. At least five orders scaling regime in $ln \, r$ is clear and $D_2 = 2.15 \pm 0.03$. 
$D_2$ increases  with $\dot \epsilon_a$ marginally until $50$. Beyond this value, we find the scaling regime shrinks. Indeed,  even for $\dot\epsilon_a=60$, $D_{2s}(r,d_E)$ increases steadily but slowly, for small length scales as is clear from Fig. \ref{Corrd90}(a).  Even when the propagation distance is not too large we find practically no scaling regime as illustrated in Fig. \ref{Corrd90}(b) for $\epsilon_a=90$. The nonconstancy of $D_{2s}(r,d_E)$ for $\dot\epsilon_a > 50$ is precisely the region of the partially propagating dislocation bands (see Fig. \ref{Band90}(a)).  

\begin{figure}
\includegraphics[height=4.5cm,width=8.0cm]{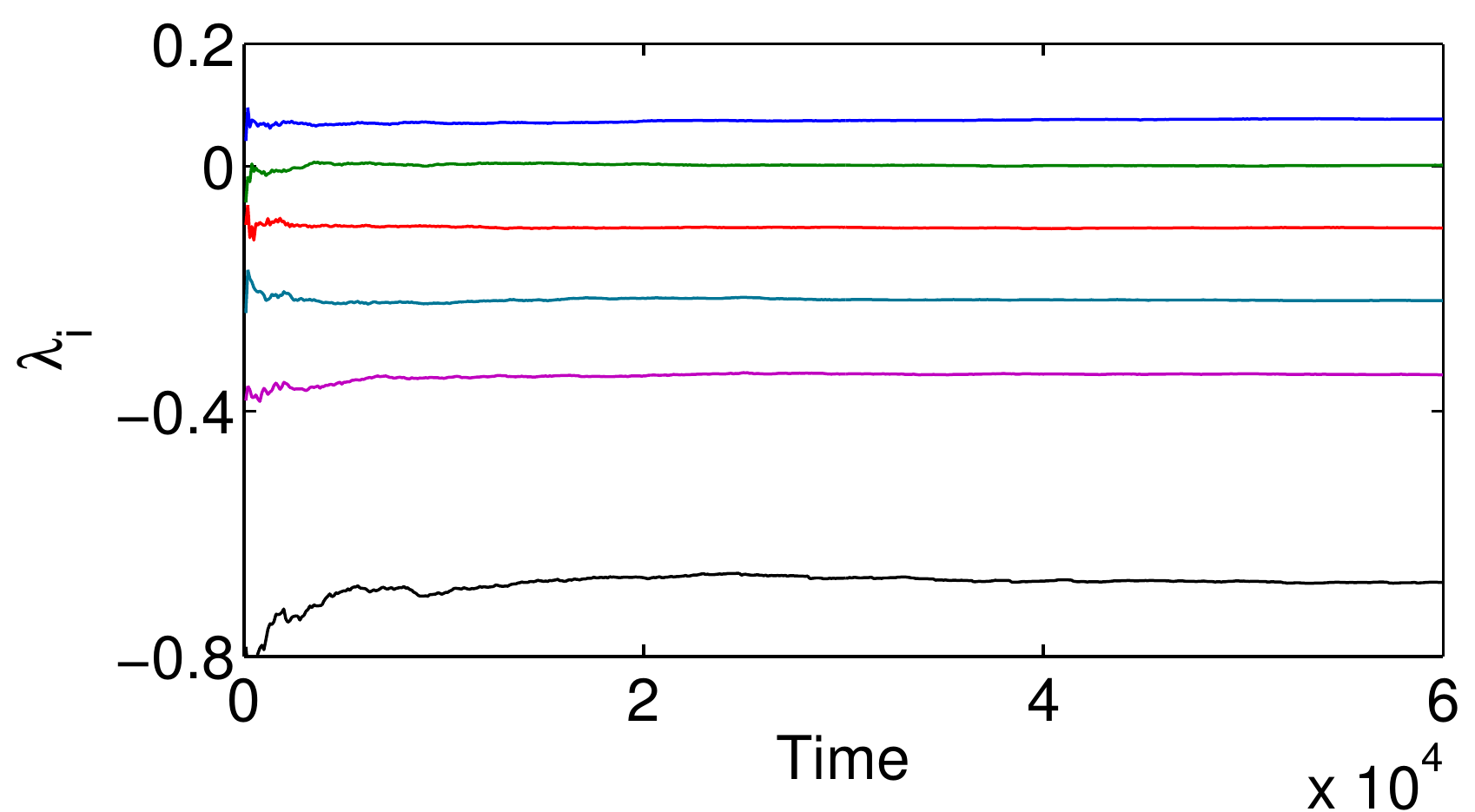}
\caption{(Color online) The Lyapunov spectrum for $d_E=6$ and $\tau=2$ for the stress-strain series for $\dot\epsilon_a=25$. The Lyapunov dimension is $2.78$. 
}
\label{Lyp25}
\end{figure}

Using the Eckmann's algorithm, we have calculated the  Lyapunov spectrum for the time series in the interval $20 < \dot\epsilon_a \le 50$. 
Fig. \ref{Lyp25} shows  the spectrum  for $\dot\epsilon_a=25$ for $d_E=6$. ( Note that $d_E=6$ corresponds to $2D_2 +1$. However, Ding {\it et al.} \cite{Ding93} have shown that it is adequate to use embedding dimension $d_E \ge D_2$ rather than using a $d_E$ larger than $2D_2+1$ as originally suggested.) The existence of a good zero exponent and a positive exponent is clear and the Lyapunov dimension $D_L = 2.78$. Similar results are obtained for $20 <\dot\epsilon_a \le 50$.   $D_L$ ranges from $2.6 - 2.8$. The minimum degrees of freedom required for a dynamical description is then four, which is also the dimension of the bare (space independent) AK model. Here, we note that as we increase the embedding dimension $d_E$ from $D_2+1$ to $2D_2 +1$, the value of the positive exponent remains nearly constant at $\approx 0.075$, the zero exponent remains close to zero ($\sim 10^{-3}$) and the first negative exponent also remains nearly constant. These three exponents are nonspurious.   For ${\dot\epsilon}_a \ge 60$, it is not meaningful to calculate the Lyapunov spectrum due to lack of convergence of  $D_{2s}(r,d_E)$.

We have also carried out surrogate (phase-randomized) data analysis of the stress-time series for $\dot\epsilon_a=25-50$. We find that $D_{2s}(r,d_E)$ increases with the embedding dimension. This provides an additional support that  the time series has features of low dimensional chaos.

\section{Summary and Conclusions}

In summary, we have demonstrated that low-dimensional chaos is detected in the stress signals for the low-strain-rate regime even as the model equations are spatiotemporally chaotic. The scaling regime for the correlation dimension shrinks beyond $\dot\epsilon_a=50$ where the tendency for propagation increases with the applied strain rate. The number of positive Lyapunov exponents and the Lyapunov dimension for the model equations scale with the system size for the low strain regime of randomly nucleated bands.  These two results  appear to be mutually inconsistent if one goes by the general belief that a scalar time series obtained from a spatially extended system should in principle contain information about the  full attractor of the system. However,  the impressive (five orders in $ln \, r$ at low applied strain rates) scaling regime for $D_2$ for the stress-time series strongly supports the low-dimensional chaotic nature of the stress signals. Interestingly, the value of $D_2$ for the stress-time series obtained from the bare (space-independent) AK model equations is also $\sim 2.25$.  Furthermore, the Lyapunov spectrum calculated from the embedded time series shows a positive and a good zero exponent, which confirms the low-d chaotic nature of the stress signal. 

On the other hand, the scaling regime for the correlation dimension shrinks for higher strain rates ($\dot\epsilon_a > 50$) once the bands acquire the propensity to propagate. In the partially propagative regime the Lyapunov dimension $D_L$ exhibits two slopes as a function of the system size $N$, one for the small sizes ($N<50$) and  another for larger sizes while we find  $D_L$ again scales with $N$ for the  fully propagating bands at high strain rates as for the low strain rates. We have shown that the two-slopes nature is due to the altered spatiotemporal patterns when the system size is reduced from those manifesting for large system sizes. 

\begin{figure}
\vbox{
\includegraphics[height=4.2cm,width=8.0cm]{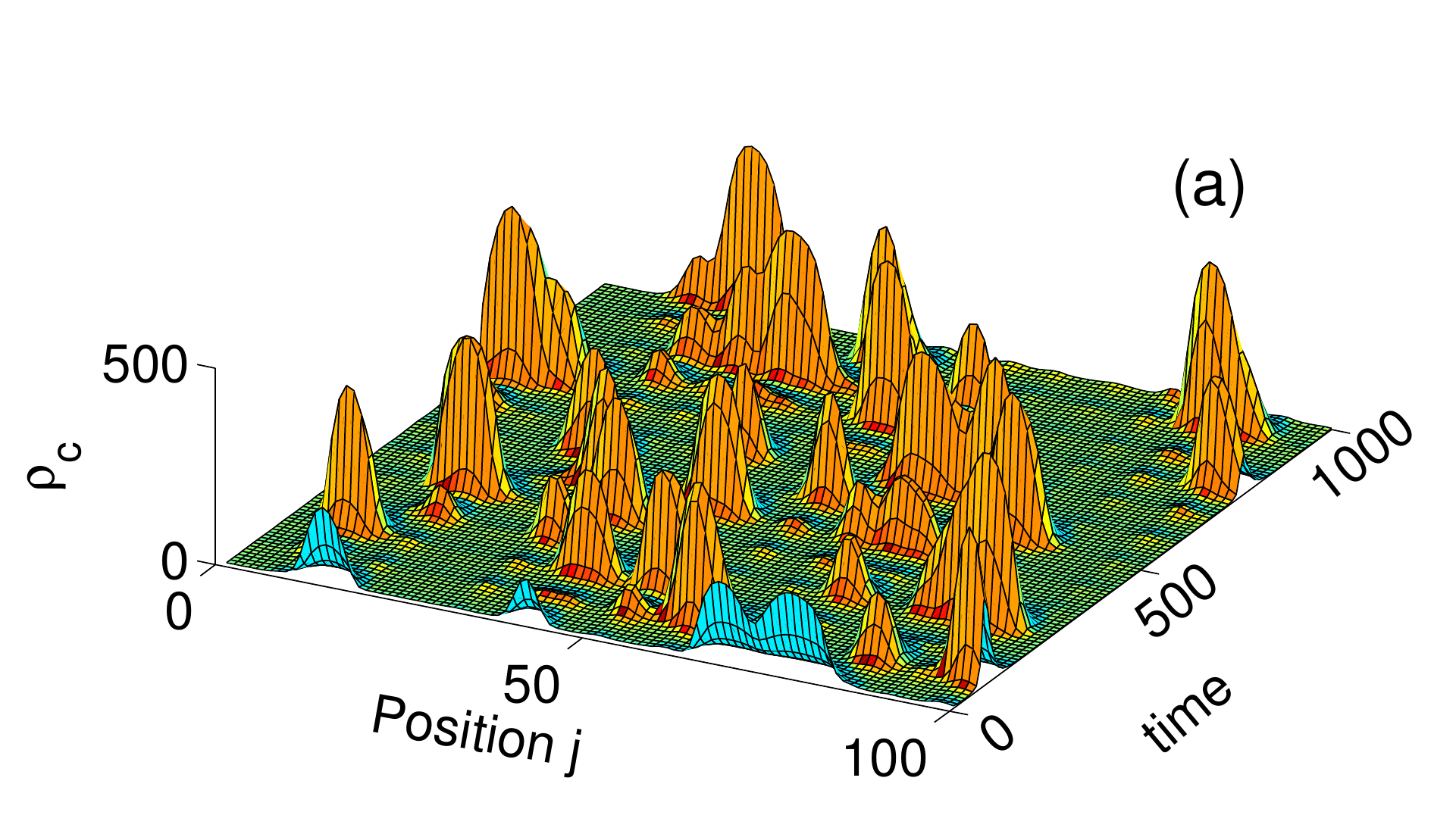}\\
\includegraphics[height=4.2cm,width=8.0cm]{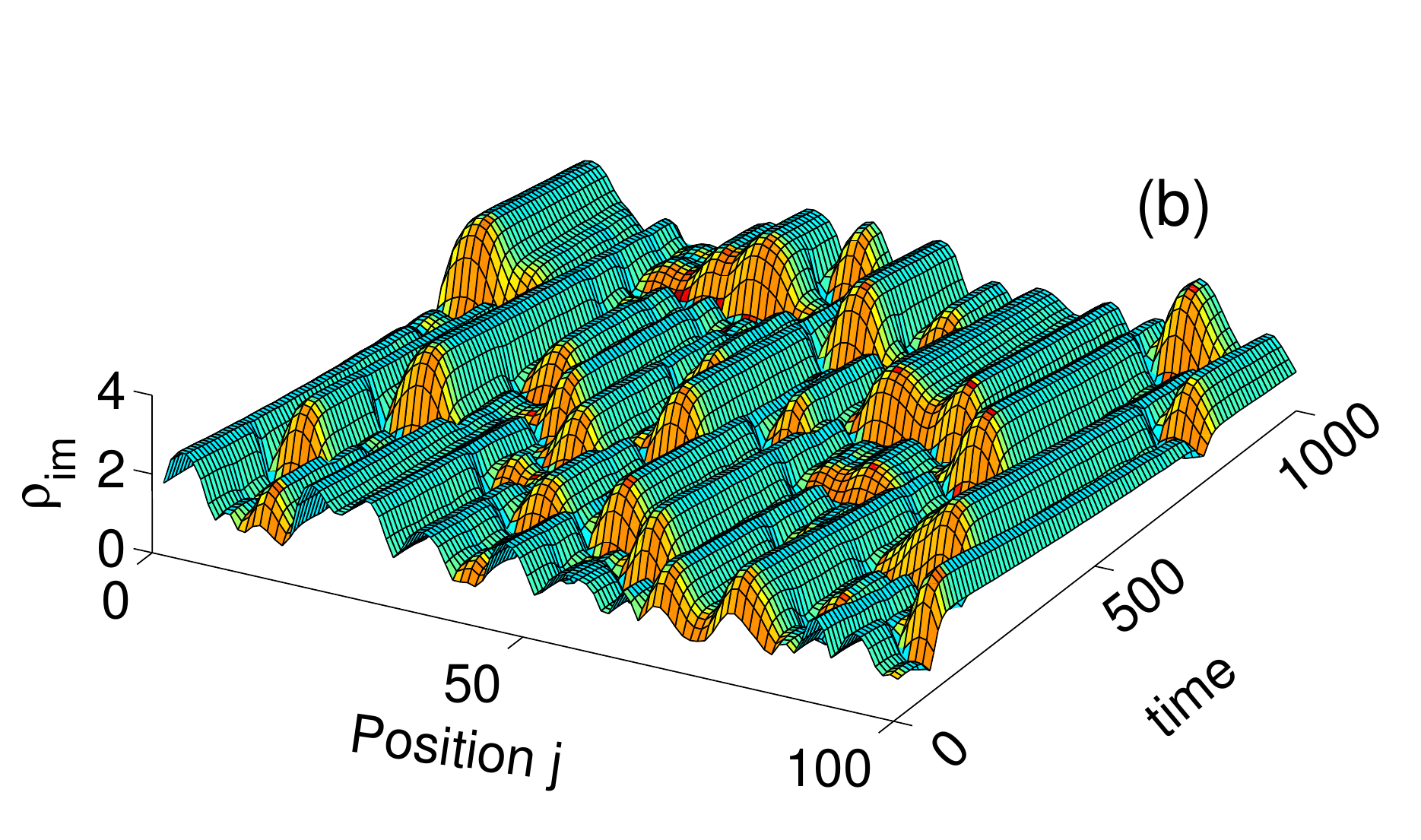}\\
\includegraphics[height=4.2cm,width=8.0cm]{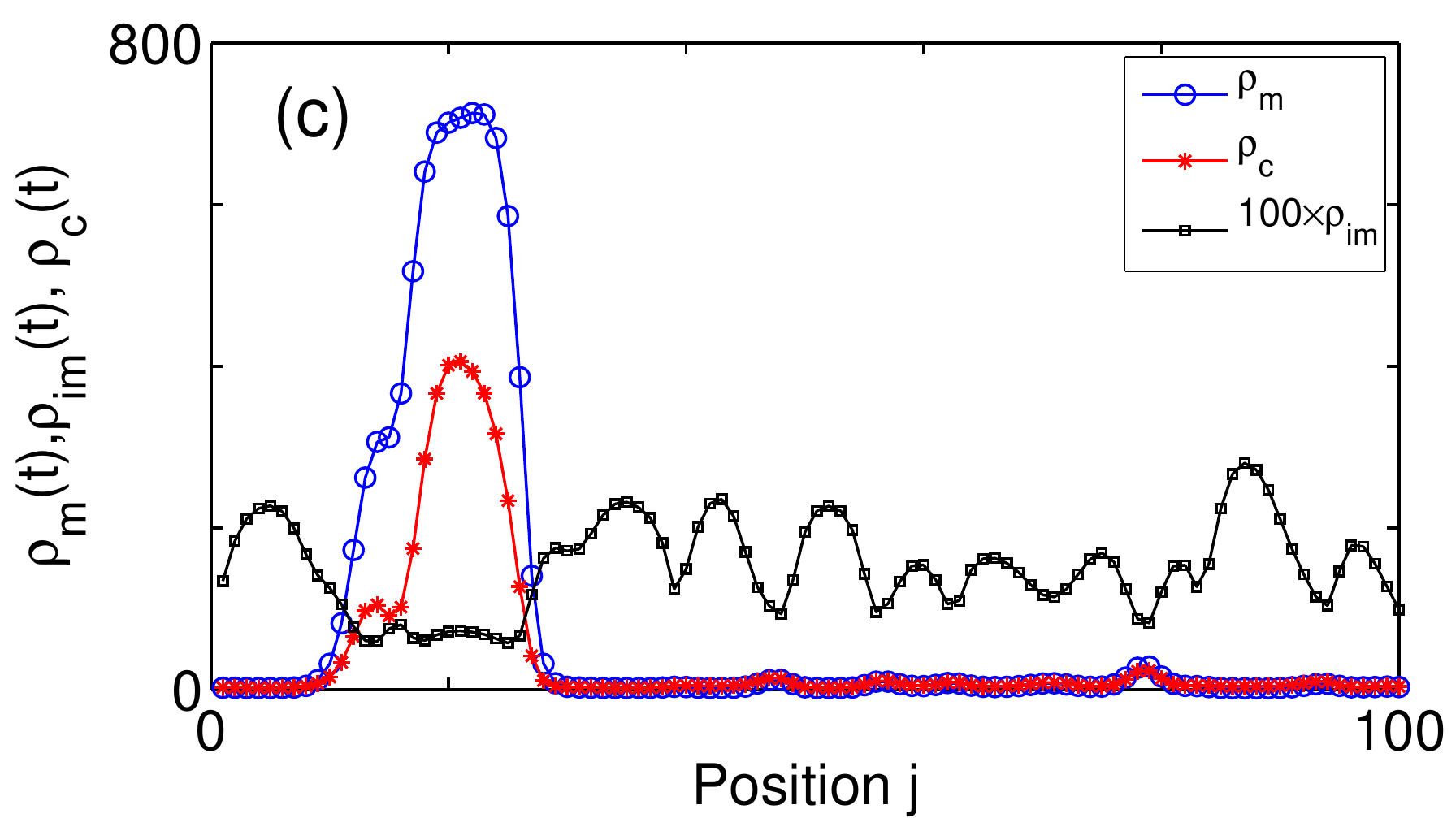}
}
\caption{(Color online)
(a) Space-time plot of $\rho_c(x,t)$ for $\dot\epsilon_a=40$ for the same time interval as for $\rho_m(x,t)$ shown in Fig. \ref{Band40}(a). (b) The corresponding space-time plot for $\rho_{im}(x,t)$.  Note that for $\rho_{im}(x,t)$ is out of phase with $\rho_m$ and $\rho_c$ for all sites. (c) A snap shot of $\rho_m(x,t),\rho_c(x,t)$ and $\rho_{im}(x,t)$ at an arbitrary  time. }
\label{Compare-c-Band40} 
\end{figure}

Having demonstrated that the stress signals have all features of low-dimensional chaos for the region of strain rates $20< \dot\epsilon_a \le 50$,   a natural question is as follows: How is the low-d chaos projected  from spatiotemporal chaotic system?   The answer lies in the nature  of spatiotemporal patterns seen at low strain rates.  Plots of $\rho_c(x,t)$ and $\rho_{im}(x,t)$ corresponding to $\rho_m(x,t)$ shown in Fig. \ref{Band40}(a) for $\epsilon_a=40$ are displayed in Figs. \ref{Compare-c-Band40}(a) and \ref{Compare-c-Band40}(b), respectively. Since the relative magnitudes of the three densities are substantially different, this feature is better captured by a snap shot representing the magnitudes of  the three densities (at an arbitrary time) as shown in Fig. \ref{Compare-c-Band40}(c). As demonstrated earlier, there is a one-to-one correspondence between the stress drops and bursts of $\rho_m$.  Furthermore, on comparing $\rho_m(x,t)$ shown Fig. \ref{Band40}(a) with $\rho_c(x,t)$ [Fig. \ref{Compare-c-Band40}(a)], it is clear that these two densities are in phase and are localized to the same spatial extent of few sites with large peak heights of $\rho_m(x,t) \sim 1000$ and $\rho_c(x,t) \sim 500$ respectively. In contrast, the range of values of $\rho_{im}(x,t)$ is two to three orders lower than $\rho_m$ and $\rho_c$ as can be seen from Fig. \ref{Compare-c-Band40}(c). Moreover, the range of $\rho_{im}$ is small with values between $0.5-3$. [Note that larger values of $\rho_{im}$ are for sites where $\rho_m$ (or $\rho_c$) is small and vice versa, as can be seen from Fig. \ref{Compare-c-Band40}(c).]  Moreover, we note that the spatial dependence in the model comes entirely from Eq. (\ref{X-eqn}).  Using these features,  we shall  now show that the spatial averaging process does project low-dimensional chaos.

To do this, we first note that  since  the contribution to the plastic strain rate [$\dot \epsilon_p=\frac{\phi^m(t)}{l}\int_0^l \rho_m(x,t) dx$] comes from the single burst, it is natural to use the space-averaged dislocation densities to illustrate the projection process.  Second, since both $\rho_m$ and  $\rho_c$ are localized to a few sites with large peak heights, they can be represented by  Gaussian-like functions. Now consider integrating Eq. (\ref{X-eqn}).  
\begin{eqnarray}
\nonumber
\int \frac{\partial \rho_m}{\partial t} dx &=& \int dx\big[-b_0  \rho_m^2 -\rho_m\rho_{im} + \rho_{im} -a\rho_m \\
&& + \phi^m \rho_m + \frac{D\phi^m(t)}{\rho_{im}} \frac{\partial^2 \rho_m}{\partial x^2}\big]
\label{Int-X}
\end{eqnarray} 
Then, define $\int \rho_m(x,t) dx/l = \bar{\rho}_m(t)$ with similar definitions for $\rho_c(x,t)$ ($\bar \rho_c(t)$) and $\rho_{im}(x,t)$ ($\bar \rho_{im}(t)$).  The linear terms in $\rho_m(x,t)$ pose no problems.  Using  a Gaussian representation  for $\rho_m(x,t)$, the first term $ \int dx \rho_m^2(x,t)/l$ in Eq. (\ref{Int-X}), can be integrated to give $ \int dx \rho_m^2(x,t)/l = \frac{l}{\bar\sigma \sqrt \pi}{\bar\rho_m^2(t)}$ where $\bar\sigma$ is the variance of the Gaussian distribution.  

Consider  evaluating $\int \rho_m(x,t)\rho_{im}(x,t) dx/l$. This involves integrating the sharply peaked $\rho_m(x,t)$ with the weight factor $\rho_{im}(x,t)$. Noting that $\rho_{im}(x,t)$ varies in a narrow range of $0.5-3$, and  $\rho_m(x,t)$ is localized to a few sites, we may approximate $\int \rho_m(x,t)\rho_{im}(x,t)/l\approx q{\bar \rho_m(t)} {\bar\rho}_{im}(t)$ where $\int \rho_{im}(x,t) dx/l = \bar\rho_{im}(t)$ with a prefactor $q$ to account for the average weight factor arising from $\rho_{im}(x,t)$ to the integral. This approximation has been verified numerically for various intervals of time. The factor $q$ is around $2$.  

Now consider evaluating the last term in Eq. \ref{Int-X}. Noting again that the range of $\rho_{im}(x,t)$ is nearly constant of the order of unity, we can replace it with  $ r \bar\rho_{im}(t)$, where $r$ is a scale factor.  Then,  using a Gaussian centered around some site, we get  
\begin{eqnarray}
\nonumber
I  &=& \int \frac{D\phi^m(t)}{\rho_{im}(x,t)} \frac{\partial^2  \rho_m(x,t)}{\partial x^2}dx \\
&& \approx  \frac{D}{r \bar\rho_{im}(t)} \phi^m(t) \int_0^l \frac{\partial^2 ( \rho_m(x,t))}{\partial x^2}dx = 0.
\end{eqnarray}
When the $\rho_m$ burst is near the boundary, this term would be nonzero. However, such events are rare and for all purposes, they may be ignored. Thus we have 
\begin{equation}
\frac{\partial {\bar\rho_m}}{\partial t} = -b'_0 {\bar\rho_m}^2 -q{\bar\rho_m}{\bar\rho}_{im} + \bar\rho_{im} -a\bar\rho_m + \phi^m \bar\rho_m.
\label{x-eqn}
\end{equation}
Here, $b_0'=  b_0\frac{l}{\bar\sigma \sqrt \pi}$.

It is straight forward to show that the other three equations can be integrated to give 
\begin{eqnarray}
\label{y-eqn}
\frac{\partial \bar\rho_{im}}{\partial t} &=& b_0'(b_0'\bar \rho_m^2 -q\bar\rho_m\bar\rho_{im} -\bar\rho_{im} + a\bar\rho_c),\\
\label{z-eqn}
\frac{\partial \bar\rho_c}{\partial t} &= &c(\bar\rho_m - \bar\rho_c ),\\
\label{phi-eqn1}
\frac{d\phi(t)}{dt}& = &d[\dot{\epsilon}_a -\bar\rho_m(t)\phi^m(t)].
\end{eqnarray}
Thus, for low strain rates, Eqs. (\ref{X-eqn})$-$(\ref{phi-eqn}) reduce to a set of  coupled ordinary differential equations for the space averaged densities given by Eqs. (\ref{x-eqn})$-$(\ref{phi-eqn1}), with renormalized co-efficients. These equations have the same form as the  bare AK model equations that have been shown to be chaotic (see Refs. \cite{GA07, Anan83}).  Indeed, the value of $D_2$ for the stress-time series obtained from the bare model turns out to be $D_2 = 2.25 \pm 0.05$ using just $30,000$ points. Clearly, the above procedure works as long as there is a one-to-one correspondence between the bursts of $\rho_m$ and stress drops. However, this correspondence breaks down even for a small extent of propagation.  Basically, during propagation, the successive bursts of $\rho_m$ ahead of the primary burst do not have any specific relation to the stress drop.

The fact that the zeroth order Fourier component captures the low-dimensional chaotic nature of the stress signal is  possibly suggestive of separation of time scales.   Here, we note that the burst time scale of $\rho_m(x,t)$ is short and the spatial averaging process maps the burst time scale to the time scale of the stress drop. Then, the one-to-one correspondence between the burst in $\rho_m$ and stress drop leads to the low-d chaotic character of the stress reflected in the constancy of $D_{2s}(r,d_E)$ for all but small length scales. However, in principle, the influence of higher  Fourier components on the zeroth Fourier component must be reflected in some form. The effect of other degrees of freedom has been represented as noise to explain the generally increasing  trend of the correlation dimension for small length scales in studies on coupled maps \cite{Olbrich98}.   Indeed, the increasing trend of $D_{2s}(r,d_E)$ for small scales  seen in Fig. \ref{Corrd25} is perhaps a reflection of the high-dimensional nature of the full system.  

Now consider the Lyapunov spectrum obtained from the two methods. The main contribution to the divergence of orbits  of the full set of model equations comes from the $\rho_m(x,t)$ and $\rho_c(x,t)$ due to the large range of values while both $\rho_{im}$ and  stress contribute minimally. In contrast, the contribution to the divergence of the orbits in the embedded  space of $\phi(t)$ comes mainly from the rapid changes occurring during stress drops.  Since the underlying mechanisms and magnitudes of the changes contributing to the Lyapunov spectrum in the two cases are very different, the  results are not necessarily  mutually {\it inconsistent}.   

We believe that projecting low-dimensional chaos from spatiotemporal chaotic dynamics should hold at least in situations where there is a one-to-one correspondence between the abrupt variation of a scalar time series (which is some kind of average over spatial degrees of freedom)  and localized excitations of the internal degrees of freedom. This could also arise due to synchronization of a certain region of spatial elements. Attempts to verify this conjecture are  underway.  Detailed investigations  are in progress  to understand the relationship between the stress signals and the spatiotemporal dynamics of the model equations, and quantifying stress signals for higher strain rates where partial propagating bands are seen.

\begin{acknowledgments}
G.A. would like to acknowledges support from  Board of Research in Nuclear Sciences (BRNS) Grant No. 2007/36/62 and support from  Indian National Science Academy for Senior Scientist Position. R.S. would acknowledges support from Council of Scientific Industrial
Research and BRNS Grant No. 2007/36/62.
\end{acknowledgments}

\end{document}